\documentclass[paper]{JHEP3}
\usepackage{times}
\usepackage{bbm, amsmath, amsthm, amssymb}
\usepackage{mathrsfs}
\usepackage[dvips]{graphicx}
\usepackage{picinpar}

\newcommand{\GeV}{\, \textrm{GeV}}

\newcommand{\myE}{\mathcal{E}}
 
\title{Full Boltzmann equations for leptogenesis including scattering}

\author{{\slshape F.~Hahn-Woernle$\,^1$, M.~Pl\"umacher$\,^1$ and Y.~Y.~Y.~Wong$\,^{2,3}$}\\
  $^1$Max-Planck-Institut f\"ur Physik, F\"ohringer Ring 6,
  D-80805 M\"unchen, Germany\\
  $^2$Theory Division, Physics Department, CERN, CH-1211 Geneva 23, Switzerland \\
  $^3$Institut f\"ur Theoretische Physik E, RWTH Aachen University, D-52056 Aachen, Germany}

\abstract{We study the evolution of a cosmological baryon asymmetry
  produced via leptogenesis by means of the full classical Boltzmann
  equations, without the assumption of kinetic equilibrium and
  including all quantum statistical factors. Beginning with the full
  mode equations, we derive the usual equations of motion for the
  right-handed neutrino number density and integrated lepton
  asymmetry, and show explicitly the impact of each assumption on
  these quantities.  For the first time, we investigate also the
  effects of scattering of the right-handed neutrino with the top
  quark to leading order in the Yukawa couplings by means of the
  full Boltzmann equations.  We find that in our full Boltzmann
  treatment the final lepton asymmetry can be suppressed by as much as
  a factor of $\sim 1.5$ in the weak wash-out regime ($K \lesssim 1$),
  compared to the usual integrated approach which assumes kinetic
  equilibrium and neglects quantum statistics.  This suppression is in
  contrast with the enhancement seen in some previous studies that
  considered only decay and inverse decay of the right-handed
  neutrino.  However, this suppression quickly decreases as we
  increase $K$.  In the strong wash-out regime ($K \gtrsim 1$), the
  full Boltzmann treatment and the integrated approach give nearly
  identical final lepton asymmetries (within 10\% of each other at
  $K>3$).  Finally, we show that the opposing effects of quantum
  statistics on decays/inverse decays and the scattering processes
  tend to reduce the net importance of scattering on leptogenesis in
  the full treatment compared to the integrated approach.}

\preprint{CERN-PH-TH/2009-107\\
  MPP-2009-85\\
  PITHA 09/15}

\begin{document}


\section{Introduction}
\label{sec:introduction}
Leptogenesis~\cite{Fukugita:1986hr} provides an attractive explanation
for the baryon asymmetry of the universe.  The ingredients of this
mechanism are simple: heavy right-handed Majorana neutrinos are added
to the standard model of particle physics, which on the one hand
explains naturally the smallness of the observed neutrino masses by
way of the see-saw mechanism~\cite{Minkowski:1977sc}.  On the other
hand, the out-of-equilibrium decay of the heavy neutrino states into
leptons and Higgs particles violates $CP$, from whence a lepton
asymmetry can be generated.  This lepton asymmetry is then partially
transformed into a baryon asymmetry by anomalous processes of the
standard model called sphalerons~\cite{Klinkhamer:1984di}.

In the recent past a huge step forward has been made towards
understanding in detail the processes of leptogenesis.  Relevant
studies include leptogenesis in a supersymmetric
context~\cite{Plumacher:1997ru}, thermal
effects~\cite{Covi:1997dr,Giudice:2003jh}, analytic formulae for the
final efficiency factor~\cite{Buchmuller:2004nz}, the role of
flavour~\cite{Abada:2006fw,Abada:2006ea,Nardi:2006fx,Blanchet:2006be},
as well as leptogenesis with $CP$ violation coming only from the
measurable low-scale Pontecorvo--Maki--Nakagawa--Sakata
matrix~\cite{Anisimov:2007mw}.  Furthermore, it has been pointed out
that the classical Boltzmann equations are insufficient to describe
the transition region between the flavoured and the unflavoured
regimes~\cite{Blanchet:2006ch}; a full quantum-mechanical description
in terms of density matrices is necessary.  On this front, the
quantum-mechanical Kadanoff--Baym equations have been investigated for
toy models in extreme out-of-equilibrium
situations~\cite{Lindner:2005kv,Lindner:2007am}.

On a different front, the classical Boltzmann equations have been
solved for the first time for single momentum
modes~\cite{Basboll:2006yx}.  As one of Sakharov's
conditions~\cite{Sakharov:1967}, departure from thermal equilibrium is
crucial for the dynamic creation of a baryon asymmetry.  In the
leptogenesis scenario, out-of-equilibrium conditions are achieved when
interactions are no longer able to maintain the momentum distribution
function of the right-handed neutrino at its equilibrium value as the
universe expands.  To simplify the calculation, this non-equilibrium
process is traditionally studied by means of the integrated Boltzmann
equations~\cite{Giudice:2003jh,Kolb:1979qa,Harvey:1990qw,Plumacher:1996kc},
whereby the equations of motion for the distribution functions of all
particle species involved are integrated over momentum such that only
the evolution of the {\it number densities} is tracked.  However, in
order for the integrated equations to be in a closed form, it is
necessary to neglect quantum statistical behaviours (e.g., Pauli
blocking) and assume kinetic equilibrium for all particle species,
including the right-handed neutrino.  For particle species with gauge
interactions these assumptions seem justifiable.  For the right-handed
neutrino however, their validity is not immediately obvious.
 
To estimate the effects of kinetic equilibrium and quantum statistics,
the Boltzmann equations for the individual momentum modes have been
solved in~\cite{Basboll:2006yx}, taking into account only the decay
and inverse decay of the right-handed neutrino within the unflavoured
framework.  More recently, the mode equations have been used to study
the effect of a pre-existing asymmetry and the soft leptogenesis
scenario, again including only decays and inverse
decays~\cite{Garayoa:2009my}.  In the present work, we extend on these
previous studies by considering also scattering processes of the
right-handed neutrino with the top quark.

The paper is organised as follows: We describe the basic set-up of our
scenario in section~\ref{sec:starting-BE}.  In
section~\ref{sec:dec-in-dec} we study the simplest scenario involving
only the decay and inverse decay of the right-handed neutrino.  In
section~\ref{sec:full-BE-scat} we include scattering processes
mediated by the Yukawa interaction of heavy neutrinos with the top
quark.  We conclude in section~\ref{sec:conclusions}.

\section{Basic set-up}
\label{sec:starting-BE}

We concentrate on the simplest case of ``vanilla-leptogenesis'', in
which a lepton asymmetry is established from the decay and scattering
of the lightest heavy right-handed neutrino $N_{1}$. We neglect the
decay of the two heavier neutrino states
$N_{2,3}$~\cite{Blanchet:2006dq}, assuming that any lepton asymmetry
produced from these decays will be efficiently washed out by the
$N_{1}$ interactions.  Therefore we will drop the subscript ``1'', and
refer to the lightest right-handed neutrino simply as $N$ in the
following.  Furthermore we will work in the one-flavour approximation,
since flavour effects do not change the kinetic consideration for the
mode equations.

Generally, the Boltzmann equation (BE) for a right-handed neutrino (RHN) 
in a Friedman--Lema\^{\i}tre--Robertson--Walker framework
can be written as
\begin{equation}
  \label{eq:BE-general}
  \frac{\partial f_{N}}{\partial t}-\vert \mathbf{p}_{N} \vert H\,\frac{\partial
    f_{N}}{\partial \vert \mathbf{p}_{N}\vert}= 
  C_{D}\left[f_{N}\right]+C_{S}\left[f_{N}\right],
\end{equation}
where $f_N$ is the phase space distribution of the RHN, $\mathbf{p}_N
$ the RHN momentum, and $H$ the Hubble parameter.  On the right-hand
side, the collision integrals $C_{D}\left[f_{N}\right]$ and
$C_{S}\left[f_{N}\right]$ encode respectively the interactions of the
RHN due to decays into leptons and Higgs ($D$) and scattering
processes via Yukawa interactions with the top quark ($S$).

The Boltzmann equation for leptons (anti-leptons) with phase space
distribution $f_l$ ($f_{\overline{l}}$) has a similar form to
Eq.~(\ref{eq:BE-general}), save for the replacements $f_N \to f_l$
($f_N \to f_{\overline{l}}$) and $\mathbf{p}_N  \to 
\mathbf{p}_l $ ($\mathbf{p}_N \to 
\mathbf{p}_{\overline{l}}$).  Since we are interested in the
asymmetry between leptons and antileptons, it is convenient to define
\begin{equation}
f_{l-\overline{l}} \equiv f_{l}- f_{\overline{l}}, 
\end{equation}
and the corresponding Boltzmann equation
\begin{equation}
  \label{eq:asym-general}
  \frac{\partial f_{l-\overline{l}}}{\partial t}-\vert \mathbf{p}_{l}
  \vert H\,\frac{\partial
    f_{l-\overline{l}}}{\partial \vert \mathbf{p}_{l} \vert}= 
  C_{D}\left[f_{l-\overline{l}}\right] 
  + C_{S}\left[f_{l-\overline{l}}\right],
\end{equation}
where $C_{D,S}\left[f_{l-\overline{l}}\right] \equiv
C_{D,S}\left[f_{l}\right] - C_{D,S}\left[f_{\overline{l}}\right]$.
Integrating $f_{l-\overline{l}}$ over the lepton phase space, i.e.,
\begin{equation}
  n_{l-\overline{l}} \equiv \frac{g_l}{(2 \pi)^3} \int d^3 p_l\, f_{l-\overline{l}},
\end{equation}
with $g_l=2$, gives us the lepton asymmetry per co-moving photon,
\begin{equation}
 \label{eq:comoving}
 N_{l-\overline{l}} \equiv \frac{n_{l-\overline{l}}}{n_{\gamma}^{\rm eq}},
 \end{equation} 
 where $n_{\gamma}^{\rm eq}= (\zeta(3)/\pi^2) g_\gamma T^3$, with
 $g_\gamma=2$, is the equilibrium photon density.

Equations~(\ref{eq:BE-general}) and (\ref{eq:asym-general}) can be
further simplified by transforming to the dimensionless coordinates
$z= M/T$ and $y_i=\vert \mathbf{p}_i\vert/T$, where $M$ is the mass of
the RHN~\cite{Kawasaki:1992kg}.  Using the relation $dT/dt = - H T$,
the differential operator $\partial_t-\vert \mathbf{p}_i \vert
H \partial_{\vert \mathbf{p}_i\vert}$ becomes $z H \partial_z$, and
consequently
\begin{equation}
  \label{eq:BE-general-2}
   \frac{\partial f_{N}(z,y)}{\partial z} = \frac{z}{H(M)}\, \left(
     C_{D}[ f_{N}(z,y)]+C_{S}[ f_{N}(z,y)] \right),
\end{equation}
and
\begin{equation}
\label{eq:asym-general-2}
\frac{\partial f_{l-\overline{l}}(z,y)}{\partial z} = 
 \frac{z}{H(M)}\, \left(
     C_{D}[ f_{l-\overline{l}}(z,y)]+C_{S}[ f_{l-\overline{l}}(z,y)] \right),
\end{equation}
with $H(M)=\sqrt{4\pi^{3}g^{\ast}/45}\left(M/M_{\rm Pl}\right)$, where
$M_{\rm Pl}=1.221 \times 10^{19} \ {\rm GeV}$ is the Planck mass, and
$g^{\ast}=106.75$ corresponds to the number of relativistic degrees of
freedom in the standard model at temperatures higher than the
electroweak scale.

The Boltzmann equations~(\ref{eq:BE-general-2}) and
(\ref{eq:asym-general-2}) encode how a lepton asymmetry is generated
and washed out in an expanding universe given some specific particle
interactions.  Here, it is useful to define an efficiency factor that
measures the amount of asymmetry that has survived the competitive
production and wash-out processes,
\begin{equation}
  \label{eq:ef-fac}
  \kappa \equiv \frac{4}{3}\,\varepsilon^{-1}\,N_{l-\overline{l}}.
\end{equation}
The quantity $\varepsilon$ quantifies the amount of $CP$ violation in
the interactions, while $N_{l-\overline{l}}$ is the lepton asymmetry
produced. In the limit of a vanishing wash-out and a thermal initial
abundance for the RHN, the efficiency factor has a final value
$\kappa_{f}=1$.

\section{Decay and inverse decay}
\label{sec:dec-in-dec}

We consider first the simplest possible scenario of thermal
leptogenesis, in which only the decay and inverse decay of the RHN
into lepton $l$ and Higgs $\Phi$ pairs contribute to the evolution of
$f_N$, i.e., we set $C_S=0$ in Eqs.~(\ref{eq:BE-general-2}) and
(\ref{eq:asym-general-2}).  The decay and inverse decay of the RHN
violate $CP$ through interference of the tree-level and the one-loop
diagrams.

The collision integral for the RHN in the decay--inverse decay
picture has the following form:
\begin{equation}
  \label{eq:be-rhn-fund}
  \begin{aligned}
    C_{D}\left[f_{N}\right] =&\frac{1}{2\,E_{N}}\,\int
    \frac{d^{3}p_{l}}{2E_{l}(2\pi)^{3}}\,\frac{d^{3}p_{\Phi}}
    {2E_{\Phi}(2\pi)^{3}}\,(2\pi)^{4}
    \,\delta^{4}\left(p_{N}-p_{l}-p_{\Phi}\right)\\ & \times
    \left[f_{\Phi}\,f_{l}\,\left(1-f_{N}\right) \, \left(\vert
        \mathcal{M}_{\Phi l\rightarrow N}\vert^{2}+\vert
        \mathcal{M}_{\Phi \overline{l}\rightarrow N}\vert^{2}\right)
    \right.\\ & \left. -f_{N}\,\left(1-f_{l}\right)\,
      \left(1+f_{\Phi}\right)\, \left(\vert \mathcal{M}_{N\rightarrow
          \Phi l}\vert^{2}+\vert \mathcal{M}_{N \rightarrow \Phi
          \overline{l} }\vert^{2}\right)\,\right],
    \end{aligned}
\end{equation}
where $E_i$ and $p_i$ are, respectively, the energy and 4-momenta of
the particle species $i$, and $\mathcal{M}_{A}$ denotes the matrix
element for the process $A$.  
At tree level, the squared matrix element summed over all
internal degrees of freedom for the decay of the RHN into a pair of
lepton and Higgs particles is given by 
\begin{equation}
  \label{eq:mat-elem-decay}
  \vert  \mathcal{M}_{N\rightarrow \Phi l}\vert^{2}= 2\,
  \frac{(m_{D}^{\dagger}m_{D})_{11}}
  {v^{2}}\,p_{l}p_{N},
\end{equation}
where $v=174 \GeV$ is the vacuum expectation value of the Higgs
particle, and the Dirac mass matrix $m_{D}$ is connected to the Yukawa
coupling matrix $\lambda_{\nu}$ via $m_{D}=v\lambda_{\nu}$.

 The
integral~(\ref{eq:be-rhn-fund}) can be readily reduced to a one
dimensional form~\cite{Kaiser:1993bt}
\begin{equation}
  \label{eq:be-rhn}
  \begin{aligned}
    C_{D}\left[f_{N}\right]  &= \frac{M \Gamma_{\rm rf}}{E_{N}p_{N}}
    \int_{(E_{N}-p_{N})/2}^{(E_{N}+p_{N})/2} dp_{\Phi} \left [
      f_{\Phi}f_{l}(1-f_{N})-f_{N}(1-f_{l})(1+f_{\Phi}) \right ],
  \end{aligned}
\end{equation}
where \begin{equation}
  \label{eq:Gamma_rf}
  \Gamma_{\rm rf}=\frac{\tilde{m}_{1}M^{2}}{8\pi v^{2}}
\end{equation}
 is the total decay rate in the RHN's rest
frame, with 
  \begin{equation}
    \label{eq:m-nu-eff}
    \tilde{m}_{1}=\frac{(m_{D}^{\dagger}m_{D})_{11}}{M}
  \end{equation}
  the {\it
    effective neutrino mass}~\cite{Plumacher:1996kc}, to be compared 
  with the {\it equilibrium neutrino mass} 
  \begin{equation}
    \label{eq:m-nu-eq}
    m_{\ast}=\frac{16\pi^{\frac{5}{2}}\sqrt{g^{\ast}}}{3\sqrt{5}}\frac{v^{2}}{M_{Pl}}. 
  \end{equation}
  The decay parameter
  \begin{equation}
    \label{eq:K}
    K\equiv\frac{\Gamma_{\rm rf}}{H(M)}=\frac{\tilde{m}_{1}}{m_{\ast}} 
  \end{equation}
  controls whether the RHN decays in equilibrium ($K>1$) or out of
  equilibrium ($K<1$).

  For leptons participating in the same decay and inverse decay
  processes, the collisional integral is given by
\begin{equation}
  \label{eq:be-lep-fund}
  \begin{aligned}
    C_{D}\left[f_{l}\right] =&\frac{1}{2\,E_{l}}\,\int
    \frac{d^{3}p_{N}}{2E_{l}(2\pi)^{3}}\,\frac{d^{3}p_{\Phi}}
    {2E_{\Phi}(2\pi)^{3}}\,(2\pi)^{4}
    \,\delta^{4}\left(p_{N}-p_{l}-p_{\Phi}\right)\\ & \times \left[
      f_{N}\,\left(1-f_{l}\right)\, \left(1+f_{\Phi}\right)\, \vert
      \mathcal{M}_{N\rightarrow \Phi l}\vert^{2} \right.\\ &-
    \hphantom{l} f_{\Phi}\,f_{l}\,\left(1-f_{N}\right) \, \vert
    \mathcal{M}_{\Phi l\rightarrow N}\vert^{2} \left. \right].
  \end{aligned}
\end{equation}
An analogous expression for the anti-leptons can be derived by
replacing $f_l \to f_{\overline{l}}$, $\mathcal{M}_{N\rightarrow \Phi
  l} \to \mathcal{M}_{N\rightarrow \Phi \overline{l}}$, and $
\mathcal{M}_{\Phi l\rightarrow N} \to \mathcal{M}_{\Phi
  \overline{l}\rightarrow N}$.  Some useful relations exist between
the matrix elements following from $CPT$-invariance
\cite{Kolb:1979qa}: 
\begin{equation}
  \label{eq:cpt}
  \begin{aligned}
    \vert \mathcal{M}_{N \rightarrow \Phi l} \vert^{2} &= \vert
    \mathcal{M}_{\Phi \overline{l} \rightarrow N} \vert^{2}=
    \vert \mathcal{M}_{0}\vert^2\, (1+\varepsilon),\\
    \vert \mathcal{M}_{N \rightarrow \Phi \overline{l}} \vert^{2} &=
    \vert \mathcal{M}_{\Phi l \rightarrow N} \vert^{2}=
     \vert \mathcal{M}_{0}\vert^2 \, (1-\varepsilon),
  \end{aligned}
\end{equation}
where $\vert \mathcal{M}_{0} \vert^2$ is the tree level matrix element
given in Eq.~(\ref{eq:mat-elem-decay}).

The collision integral~(\ref{eq:be-lep-fund}) suffers from the problem
that a lepton asymmetry is produced even in thermal equilibrium.  This
can be remedied by including contributions from the resonant part of
the $\Delta L=2$ scattering process $l\Phi\leftrightarrow
\overline{l}\Phi$~\cite{Giudice:2003jh,Kolb:1979qa}. We implement this
remedy following the method developed in~\cite{Buchmuller:2004nz}, and
add to the collision integral~(\ref{eq:be-lep-fund}) the term
\begin{equation}
\label{eq:add}
f_\Phi f_{\overline{l}} \ (1-f_N)  \vert \mathcal{M}_{\Phi
  \overline{l} \rightarrow N} \vert_{\rm sub}^{2}
- f_\Phi f_l \ (1-f_N)  \vert \mathcal{M}_{\Phi l \rightarrow N} \vert_{\rm sub}^{2},
\end{equation}
where 
\begin{equation}
  \label{eq:res-sub}
  \begin{aligned}
    \vert \mathcal{M}_{\Phi \overline{l} \rightarrow N} \vert_{\rm
      sub}^{2} &= \vert \mathcal{M}
    \vert_{\Delta L=2}^{2}-\varepsilon\,  \vert \mathcal{M}_{0}\vert^2, \\
    \vert \mathcal{M}_{\Phi l \rightarrow N} \vert_{\rm sub}^{2} &=
    \vert \mathcal{M} \vert_{\Delta L=2}^{2}+\varepsilon\, \vert
    \mathcal{M}_{0}\vert^2,
  \end{aligned}
\end{equation}
where $ \vert \mathcal{M} \vert_{\Delta L=2}^{2}$ is negligible for
$M\ll 10^{14}
\GeV$~\cite{Buchmuller:2002rq}.\footnote{Reference~\cite{Garayoa:2009my}
  includes terms in addition to Eq.~(\ref{eq:add}) in order to avoid
  asymmetry production in thermal equilibrium.  However, the same
  analysis also shows that the quantitative difference between this
  and our approach is negligible.}

In the following subsections, we review first the derivation of the
conventional integrated Boltzmann equations, which neglects quantum
statistics and assumes kinetic equilibrium for the RHN.  We then
remove step by step these assumptions, in order to examine their
effects on the efficiency factor $\kappa$.  The scenarios to be
examined and their associated assumptions are summarised in
table~\ref{tab:dec}.

\begin{table}[t]
  \centering
  \caption{Scenarios considered in the decay/inverse decay picture and
    their associated assumptions.  Case D1 corresponds to the conventional
    integrated Boltzmann approach, while Case D4 was previously investigated
    by Basb{\o}ll and Hannestad~\cite{Basboll:2006yx}.}
  \label{tab:dec}
  \begin{tabular}[centered]{|c|c|c|c|c}
    \hline  & Assumption of kinetic equilibrium & Including quantum statistics & Section\\ 
    \hline Case D1 & Yes & No  & \ref{subsec:conv-BE-d} \\
    \hline Case D2 &  No  & No & \ref{subsec:drop-ke-d}\\
    \hline Case D3 & Yes  & Yes & \ref{sec:BE-stat-fac}\\
    \hline Case D4 & No & Yes& \ref{subsec:full-be-d}\\
    \hline
  \end{tabular}
\end{table}

\subsection{Case D1: Integrated Boltzmann equations}
\label{subsec:conv-BE-d}

In the integrated approach conventionally used in the
literature~\cite{Giudice:2003jh,Buchmuller:2004nz,Kolb:1979qa}, the
time evolution of number densities $n_i$ are tracked in favour of the
phase space distributions $f_i$.  This is achieved by integrating the
Boltzmann equations~(\ref{eq:BE-general-2})
and~(\ref{eq:asym-general-2}) over momentum.  However, the integrated
equations have no closed forms unless we make certain simplifying
assumptions: First, we neglect factors stemming from Pauli blocking
for fermions and induced emission for bosons, i.e., we approximate
$1\pm f_{i} \approx 1$~\cite{Kolb:1979qa}.  Second, all standard model
particles are taken to be in thermal equilibrium due to their gauge
interactions and their distribution functions approximated by a
Maxwell--Boltzmann distribution, $f_{i}^{\rm eq}=e^{-E_{i}/T}$.

With these assumptions and using energy conservation, we find
\begin{equation}
  \label{eq:en-cons}
  f_{\Phi}f_{l}=e^{-\left(E_{\Phi}+E_{l}\right)/T}=e^{-E_{N}/T}=f_{N}^{\rm eq},
\end{equation}
so that the collision integral~(\ref{eq:be-rhn}) simplifies to
\begin{equation}
  \label{eq:cNn-1}
     C_{D}[f_N]=\frac{M
      \Gamma_{\rm rf}}{E_{N}\,p_{N}}
    \int_{(E_{N}-p_{N})/2}^{(E_{N}+p_{N})/2} dp_{\Phi} \left [
      f_{N}^{\rm eq}-f_{N }\right ].
 \end{equation}
 Integrating (\ref{eq:cNn-1}) over $p_{\Phi}$ and inserting into
 Eq.~(\ref{eq:BE-general-2}), the BE for the RHN distribution function
 becomes
\begin{equation}
  \label{eq:BE-fN-1}
  \frac{\partial f_{N}}{\partial z}  = \frac{z\, \Gamma_{\rm rf}\,M
  }{H(M)\,E_{N}}
\left(f_{N}^{\rm eq}-f_{N}\right).
\end{equation}
To make further inroads, we assume kinetic equilibrium holds for the
RHN, i.e., its distribution function $f_N$ can be expressed as
$f_{N}/f_{N}^{\rm eq} \approx n_{N}/n_{N}^{\rm eq}$, where $n_N$ is
the RHN number density. Then one can easily integrate
Eq.~(\ref{eq:BE-fN-1}) over the RHN phase space to obtain
\begin{equation}
  \label{eq:BE-NN-1}
  \frac{\partial n_{N}}{\partial
    z}=  z\,K\, \left\langle
  \frac{M}{E_{N}}\right\rangle \left(n_{N}^{\rm eq}-n_{N}\right),
\end{equation}
where $K\equiv\Gamma_{\rm rf}/H(M)$ (cf.\ Eq.~(\ref{eq:K})), and
$\Gamma_{\rm rf} \langle M/E_{N}\rangle \equiv (\Gamma_{\rm
  rf}/n_N^{\rm eq}) \int d^3 p_N /(2\pi)^{3} \;f_N^{\rm eq}\ (M/E_N)$
is the thermal average of the decay rate~\cite{Kolb:1979qa}. The
thermally averaged dilation factor is given by the ratio of the
modified Bessel functions of the second kind of first and second order, $ \langle
M/E_{N}\rangle= K_{1}(z)/K_{2}(z)$.

Dividing Eq.~(\ref{eq:BE-NN-1}) by the equilibrium photon density
$n_{\gamma}^{\rm eq}$, we obtain the Boltzmann equation for the
quantity $N_{N} \equiv n_{N}/n_{\gamma}^{\rm
  eq}$~\cite{Buchmuller:2004nz},
\begin{equation}
  \label{eq:BE-Nn-2}
  \frac{\partial N_{N}}{\partial z}=-\,D\,\left(N_{N}-N_{N}^{\rm eq}\right),
\end{equation}
with 
\begin{equation}
D \equiv z\,K\,\left \langle \frac{M}{E_{N}} \right \rangle,
\end{equation} 
and 
\begin{equation}
\label{eq:neq}
N^{\rm eq}_{N}(z)=\frac{3}{8} \,z^{2}K_{2}\,(z). 
\end{equation}
Here, an inconsistency in the integrated approach is visible: all
particles, i.e., $N$, $l$ and $\Phi$, are assumed to follow the
Maxwell--Boltzmann distribution function. However, when calculating
$N^{\rm eq}_N$, we must use a Fermi--Dirac distribution for the RHN,
$n_{N}^{\rm eq}=[3\,\zeta(3)\,g_{N}TM^{2}/(8\pi^{2})]\,K_{2}(z)$, with
$\zeta(3)\approx 1.202$ and $g_{N}=2$, in order to reproduce a
realistic equilibrium RHN to photon density ratio.  This leads to an
extra prefactor $(3/4) \zeta(3)$ in the definition of $N_{N}^{\rm eq}$
compared to a strictly Maxwell--Boltzmann approach.

For the lepton asymmetry, Eqs.~(\ref{eq:asym-general-2}),
(\ref{eq:be-lep-fund}) and (\ref{eq:add}) combine to give the BE for
the lepton distribution functions,
 \begin{equation}
  \label{eq:BE-fa-ke-1a}
  \frac{\partial f_{l-\overline{l}}}{\partial z}
  =-\,\frac{z^{2}\,K} {2\,y^{2}_{l}}
  \,\int_{\frac{z^{2}-4y^{2}_{l}}{4y_{l}}}^{\infty}\,dy_{N}\, \frac{y_{N}}{\myE_{N}}\, \left[
    f_{\Phi}\,f_{l-{\overline{l}}} - 2\,\varepsilon
    \left(f_{N}-f_{N}^{\rm eq}\right) \right],
 \end{equation}
 where $\myE_{N}\equiv E_{N}/T$.  Using energy conservation and 
 assuming kinetic equilibrium for the RHN, Eq.~(\ref{eq:BE-fa-ke-1a}) 
 can be integrated over $y_{N}$ to give~\cite{Basboll:2006yx}
\begin{equation}
  \label{eq:BE-fa-1}
    \frac{\partial f_{l-\overline{l}}}{\partial z}=
    -\frac{z^{2}\,K}{2\,y_{l}^{2}}
    \,e^{-\frac{z^{2}+4y_{l}^{2}}{4y_{l}}} \,
    \left[e^{y_{l}}\,f_{l-{\overline{l}}} - 2\,\varepsilon\left(
        \frac{n_{N}-n_{N}^{\rm eq}}{n_{N}^{\rm eq}} \right) \right].
\end{equation}
We further assume that both kinetic chemical equilibrium prevails for
the leptons such that
\begin{equation}
\label{eq:chemical}
\begin{aligned}
f_{l-{\overline{l}}}^{\rm eq} &= e^{-(E_l-\mu)/T}-e^{-(E_l+\mu)/T}
\approx 2\,(\mu/T)\,e^{-E_{l}/T}, \\
n_{l-\overline{l}}^{\rm eq} &\approx 2\, (\mu/T) \,n_{l}^{\rm eq},\\
f_{l-\overline{l}} &\approx \frac{n_{l-\overline{l}}}{n_l^{\rm eq}} e^{-y_l},
\end{aligned}
\end{equation}
with chemical potential $\mu \ll 1$ and $n_{l}^{\rm eq}$ the lepton
equilibrium number density. 
Thus, integrating over the lepton phase space,
we obtain the equation of motion for the number density
\begin{equation}
  \label{eq:BE-Na-1}
     \frac{\partial n_{l-\overline{l}}}{\partial
      z}= -\frac{z^{3}\,K\,T^{3}}{2\,\pi^{2}} \,K_{1}(z) \left[
      \frac{n_{l-\overline{l}}}{n_{l}^{\rm eq}} - 2\,\varepsilon\left(
        \frac{n_{N}-n_{N}^{\rm eq}}{n_{N}^{\rm eq}} \right) \right],
 \end{equation}
 where $K_1(z)$ is the modified Bessel function of first
 kind. Following~\cite{Buchmuller:2004nz} we rewrite
 Eq.~(\ref{eq:BE-Na-1}) in terms of the lepton asymmetry per co-moving
 photon ,
\begin{equation}
  \label{eq:BE-Na-2}
  \frac{\partial N_{l-\overline{l}}}{\partial z}=
  -W_{ID}\,N_{l-\overline{l}}+\varepsilon\,D\left(N_{N}-N_{N}^{\rm eq}\right),
\end{equation}
where 
\begin{equation}
\label{eq:wid}
W_{ID} \equiv \frac{1}{4} \,Kz^{3}\,K_{1}(z) = \frac{1}{2}\,D\,\frac{N^{\rm eq}_{N}}{N^{\rm eq}_{l}}
\end{equation} 
quantifies the strength of the wash-out due to inverse decays, and
$N_l^{\rm eq}=3/4$.  Note that, as with the RHN, when evaluating
$N_l^{\rm eq}$ it is necessary to use a Fermi--Dirac distribution for
the leptons, $n_{l}^{\rm eq}=(3/4)\, (\zeta(3)/\pi^2) g_l T^3$, with
$g_{l}=2$, to ensure a realistic lepton to photon density ratio.

\begin{figure}[t]
  \centering
  \includegraphics[width=1.0\textwidth]{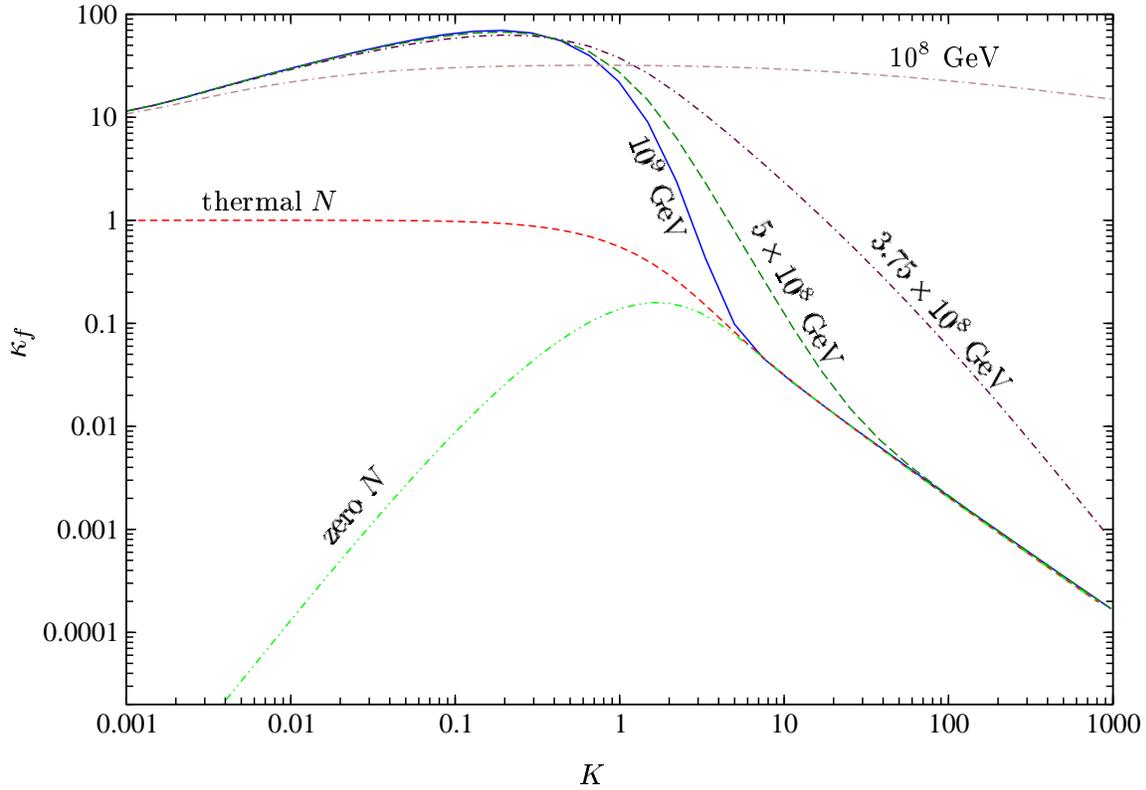}
  \caption{Final efficiency factor for different scenarios of thermal
    and non-thermal leptogenesis. Shown are $\kappa_{f}$ for a thermal
    (dashed/red), a vanishing (dot-dot-dash/light green), and several
    cases of dominant initial RHN abundance. A dominant initial
    abundance is realized if a scalar field responsible for inflation
    decays exclusively into the RHN, which then dominates the energy
    density of the universe. The coupling strength between the scalar
    field and the RHN can be connected to an energy scale: $10^{9}
    \GeV$ (solid/blue), $5\times 10^{8} \GeV$ (dashed/dark green),
    $3.75\times 10^{8} \GeV$ (dot-dash/purple) and $10^{8} \GeV$
    (dot-dash-dash/taupe).}
  \label{fig:k_f_conv}
\end{figure}

Figure~\ref{fig:k_f_conv} shows the final efficiency factor $\kappa_f$,
defined in Eq.~(\ref{eq:ef-fac}), as a function of $K$ for several
different initial RHN abundances~\cite{HahnWoernle:2008pq}.

\subsection{Case D2: Dropping the assumption of kinetic equilibrium}
\label{subsec:drop-ke-d}

Since the RHN is very heavy---its mass scale corresponds to the
temperature of the thermal bath during the period of leptogenesis---it
is not {\it a priori} obvious that decays and inverse decays would
occur fast enough to establish kinetic equilibrium.  Thus the
assumption of kinetic equilibrium for the RHN might lead to sizable
deviations from an exact treatment.  In this section we drop this
assumption in our calculation of the efficiency factor.  We retain
however our other assumptions: that all equilibrium distribution
functions are of the Maxwell--Boltzmann form, and quantum statistical
factors are negligible.

Dropping the assumption of kinetic equilibrium for the RHN means that
it is now necessary to solve Eq.~(\ref{eq:BE-fN-1}), rewritten here as
\begin{equation}
  \label{eq:BE-fN-ke-1}
  \frac{\partial f_{N}}{\partial z} =\frac{z^{2}\,K}{\myE_{N}} \,
  \left(e^{-\myE_{N}}-f_{N}\right),
\end{equation}
individually for all possible values of the dimensionless RHN energy
$\myE_{N}$.  For the calculation of the lepton asymmetry, the relevant
equation is Eq.~(\ref{eq:BE-fa-ke-1a}) which we reproduce here:
\begin{equation}
  \label{eq:BE-fa-ke-1}
  \frac{\partial f_{l-\overline{l}}}{\partial z}
  =-\,\frac{z^{2}\,K} {2\,y^{2}_{l}}
  \,\int_{\frac{z^{2}-4y^{2}_{l}}{4y_{l}}}^{\infty}\,dy_{N}\, \frac{y_{N}}{\myE_{N}}\, \left[
    f_{\Phi}\,f_{l-{\overline{l}}} - 2\,\varepsilon
    \left(f_{N}-f_{N}^{\rm eq}\right) \right].
 \end{equation}
 Again, this equation must be solved for all possible values of the
 lepton momentum $y_l$, and the resulting $f_{l-\overline{l}}(y_l)$
 summed according to Eq.~(\ref{eq:comoving}) to give
 $N_{l-\overline{l}}$.  Alternatively, using energy conservation and
 assuming kinetic and chemical equilibrium for the standard model
 particles, we can integrate Eq.~(\ref{eq:BE-fa-ke-1}) over the lepton
 phase space to obtain a single equation of motion for
 $N_{l-\overline{l}}$,
\begin{equation}
  \label{eq:BE-Na-ke-2}
  \frac{\partial
    N_{l-\overline{l}}}{\partial z}=-\frac{z^{2}\,K}{4}\,
  \int_{0}^{\infty}dy_{l}
  \int_{\frac{z^{2}-4y_{l}^{2}}{4y_{l}}}^{\infty}\,dy_{N}\,\frac{y_{N}}{\myE_{N}}\, \left[
    N_{l-\overline{l}} \, f_{N}^{\rm eq}
    -2\,\varepsilon \, \left(f_{N}-f_{N}^{\rm eq}\right)\right].
\end{equation}
 We find the second approach to yield more stable results.

\subsection{Case D3: Boltzmann equations with quantum statistical factors}
\label{sec:BE-stat-fac}

In Case D3 we reinstate Pauli blocking factors for fermions
and factors due to induced emission for bosons, but adopt again the
assumption of kinetic equilibrium for the RHN.  Consistency requires that
we use the Fermi--Dirac and the Bose--Einstein distribution
functions respectively for fermions and bosons in thermal equilibrium,
instead of the classical Maxwell--Boltzmann distribution function.
  
With these assumptions in mind,  we 
integrate the collision integral~(\ref{eq:be-rhn}) over $p_{\Phi}$ to obtain
the BE for the RHN,
\begin{equation}
  \label{eq:BE-fN-qs-1}
  \frac{\partial f_{N}}{\partial z} =
  \frac{z^{2}\,K}{\myE_{N}\,y_{N}} \, \frac{n_{N}-n_{N}^{\rm eq}}{n_{N}^{\rm eq}}
  \, \frac{1}{e^{\myE_{N}}+1} \,\log \left[ \frac{ \sinh
      \left(\left( \myE_{N}-y_{N}\right)/2\right)}{ \sinh
      \left( \left(\myE_{N}+y_{N}\right)/2\right)} \right],
\end{equation}
where we have used $f_N/f_N^{\rm eq} = (1+e^{\myE_N}) f_N \approx
n_N/n_N^{\rm eq}$.  Integrating over the RHN phase space and
normalising to the photon number density yields
\begin{equation}
  \label{eq:BE-Nn-qs-1}
  \frac{\partial N_{N}}{\partial z}=
  \frac{K}{K_{2}(z)}\,\left(N_{N}-N_{N}^{\rm eq}\right)\,
  \int_{0}^{\infty} \, dy_{N} \,\frac{y_{N}}{\myE_{N}}\,
  \frac{1}{e^{\myE_{N}}+1} \,\log \left[ \frac{ \sinh
      \left(\left( \myE_{N}-y_{N}\right)/2\right)}{ \sinh
      \left( \left(\myE_{N}+y_{N}\right)/2\right)} \right].
\end{equation}
We note that the integral over the RHN phase space has no
simple analytic form.  Therefore it remains necessary to 
perform the integration numerically.

The BE for the lepton asymmetry including all quantum statistical factors
and assuming kinetic equilibrium for all particle species has the following
form:
\begin{equation}
  \label{eq:BE-fa-full-1}
  \frac{\partial f_{l-\overline{l}}}{\partial z} =
  -\frac{z^{2}\,K}{2y^{2}_{l}} \int_{\frac{z^{2}-4y^{2}_{l}}{4y_{l}}
  }^{\infty} dy_{N}\,\frac{y_{N}}{\myE_{N}} \left [ (f_{\Phi}+\frac{n_{N}}{n_{N}^{\rm eq}}\,
  f_{N}^{\rm eq})(f_{l-{\overline{l}}} +
    \varepsilon F^{+}) - 2 \varepsilon\, \frac{n_{N}}{n_{N}^{\rm eq}}\,f_{N}^{\rm eq}(1+f_{\Phi}) \right ],
 \end{equation}
where 
$F^{+} \equiv f_{l}+f_{\overline{l}}\approx 2\,f_{l}^{\rm eq}$.
After integrating over the lepton phase space and normalising to the
photon number density we arrive at
\begin{equation}
\begin{aligned}
  \label{eq:BE-Na-ke-1}
  \frac{\partial N_{l-\overline{l}}}{\partial z} = &-
  \frac{z^{2}\,K}{4} \int_{0}^{\infty}
  dy_{l}   \\
  & \times \int_{\frac{z^{2}-4y_{l}^{2}}{4y_{l}}}^{\infty} dy_{N}
  \frac{y_{N}} {\myE_{N}} \left [ (f
    _{\Phi}+\frac{N_{N}}{N_{N}^{\rm eq}}\,f_{N}^{\rm eq})\,\left(\frac{4}{3}\,N_{l-\overline{l}}
      + 2\, \varepsilon \right)\,f_{l}^{\rm eq} - 2 \varepsilon \,
    \frac{N_{N}}{N_{N}^{\rm eq}}\,f_{N}^{\rm eq}\,\left(1+f_{\Phi}\right)
  \right ],
  \end{aligned}
\end{equation}
with $N_N^{\rm eq}$ given in Eq.~(\ref{eq:neq}).

\subsection{Case D4: Complete mode equations}
\label{subsec:full-be-d}

Here we include all statistical factors and make no 
assumption of kinetic equilibrium for the RHN.  Integrating Eq.~(\ref{eq:be-rhn}) 
over $p_\Phi$ gives the BE for the RHN,
\begin{equation}
  \label{eq:BE-full-f-1}
  \frac{\partial f_{N}}{\partial z}=
  K \frac{z^{2}}{\myE_{N}\, y_{N}} 
  \frac{-1 + f_{N} + e^{\myE_{N}} f_{N}}{e^{\myE_{N}}+1} \, \log
  \left [\frac{ \sinh((\myE_{N}-y_{N})/2)}{\sinh((\myE_{N}+y_{N})/2)} \right ].
\end{equation}
The equation for the lepton asymmetry in this case is similar 
to Eq.~(\ref{eq:BE-fa-full-1}), except we do not
assume kinetic equilibrium for the RHN,
  \begin{equation}
  \label{eq:BE-fa-full-1a}
  \frac{\partial f_{l-\overline{l}}}{\partial z} =
  -\frac{z^{2}\,K}{2y^{2}_{l}} \int_{\frac{z^{2}-4y^{2}_{l}}{4y_{l}}
  }^{\infty} dy_{N}\,\frac{y_{N}}{\myE_{N}} \left [ (f_{\Phi}+f_{N})(f_{l-{\overline{l}}} +
    \varepsilon F^{+}) - 2 \varepsilon\,f_{N}\,(1+f_{\Phi}) \right ].
 \end{equation}
 Integrating over the RHN momentum
 yields
\begin{equation}
\begin{aligned}
  \label{eq:BE-Na-full-1}
  \frac{\partial N_{l-\overline{l}}}{\partial z} = &-
  \frac{z^{2}\,K}{4} \int_{0}^{\infty}
  dy_{l}   \\
  & \times \int_{\frac{z^{2}-4y_{l}^{2}}{4y_{l}}}^{\infty} dy_{N}
  \frac{y_{N}} {\myE_{N}} \left [ (f
    _{\Phi}+f_{N})\,\left(\frac{4}{3}\,N_{l-\overline{l}}
      + 2\, \varepsilon \right)\,f_{l}^{\rm eq} - 2 \varepsilon \,
   f_{N}\,\left(1+f_{\Phi}\right)
  \right ],
  \end{aligned}
\end{equation}
where we have assumed, as usual, thermal equilibrium for the standard
model particles.

\subsection{Results and discussions\label{sec:results}}

\subsubsection{Right-handed neutrino}

\begin{figure}[t]
  \centering
  \includegraphics[width=1.0\textwidth]{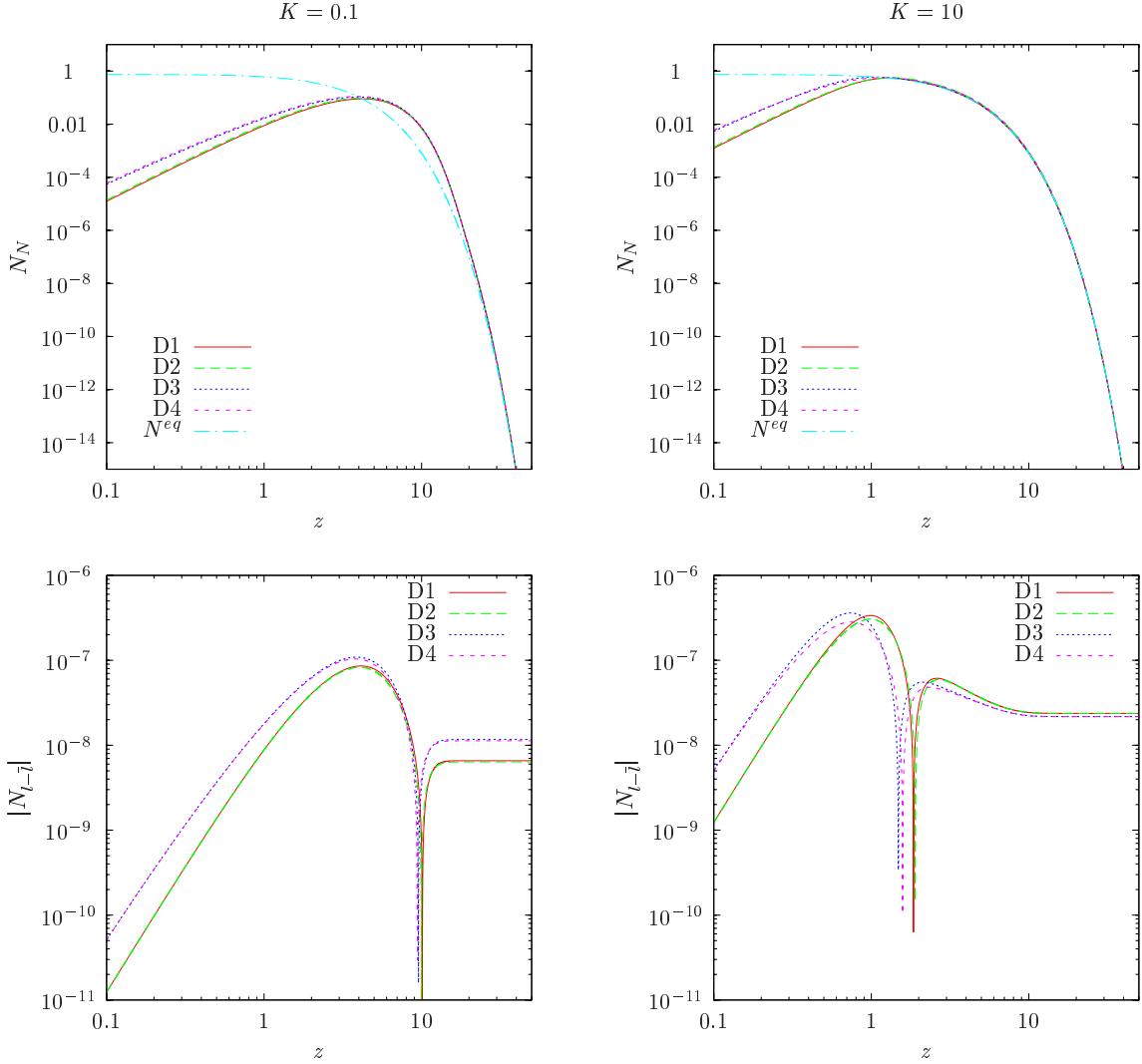} 
  \caption{Time evolution of the comoving RHN number density $N_{N}$
    and of the absolute value of the lepton asymmetry
    $N_{l-\overline{l}}$, assuming two different coupling strengths
    $K=0.1, 10$, and $\varepsilon=10^{-6}$. The four scenarios within
    the decay--inverse decay only framework are shown: Solid/red line
    denotes Case D1, long dashed/green D2, dotted/blue D3,
    and short dashed/magenta D4.  See table~\ref{tab:dec} for a
    short summary of each scenario.  For reference, we also indicate 
  the equilibrium RHN number density $N_N^{\rm eq}$ in dot-dash/cyan.}
  \label{fig:NBL-N_dec}
\end{figure}

Figure~\ref{fig:NBL-N_dec} shows the time evolution of the comoving
number densities of the RHN for the four different cases
described above, assuming a vanishing initial RHN abundance. We
have picked two values for the decay parameter: (i) $K=0.1$, lying in
the weak wash-out regime, is shown on the left hand side, and (ii)
$K=10$, lying in the strong wash-out regime, is shown on the right
hand side.   For reference, we also plot the time evolution of the 
RHN equilibrium number density $N_{N}^{\rm eq}$.

The general behaviour of the RHN abundance evolution is similar for
all four cases. In the weak wash-out regime, there is a net production
of RHN by inverse decays at high temperatures $z<1$.  At $z\sim 4$,
the RHN abundance overshoots the equilibrium density and continues to
grow until $z\sim 5$, when a net destruction of RHN by decays into
$l\Phi$ pairs begins to push its abundance slowly back down to the
equilibrium value.  Equilibrium is reached finally at $z\sim 20$,
beyond which the RHN abundance falls off exponentially with $z$, as
expected for all non-relativistic particle species in thermal
equilibrium.  Contrastingly, in the strong wash-out regime, the
stronger coupling brings the RHN abundance to its equilibrium value
already at $z\sim1$.  Its subsequent evolution is then simply governed
by equilibrium statistics: at $z \sim 4$ the RHN becomes
nonrelativistic and hence its abundance is suppressed by $\exp(-M/T)$.

In both the weak and strong wash-out regimes, the difference between
Cases D1 and D2, which exclude quantum statistical factors, and their
counterparts Cases D3 and D4, which include quantum statistics, is
most visible at $z \lesssim 1$.  The RHN abundance is almost an
  order of magnitude larger in the latter two cases than in the
former.  This is because during the high temperature RHN production
phase, using the correct Bose--Einstein equilibrium distribution
function for the Higgs boson $f_\Phi$ substantially enlarges the phase
space available for the inverse decay process $\Phi l \to N$ at low
$E_\Phi$.  This effect is far stronger than the phase space
restriction due to Pauli blocking by the final state RHN, as can be
seen from the phase space factors in Eq.~(\ref{eq:be-rhn}).  As the
temperature drops and the RHN becomes nonrelativistic, the effects of
quantum statistics also diminish, since kinematics now prevents the
low energy $\Phi$ and $l$ states from contributing to the collision
integrals.

Interestingly, the assumption of kinetic equilibrium leads to no
visible effects in either the weak or strong wash-out regime.
Comparing Cases D1 and D2 (both assume Maxwell--Boltzmann statistics),
their RHN abundances are virtually identical.  The same is true for
Cases D3 and D4, which include quantum statistical factors.

\subsubsection{Lepton asymmetry}

\begin{figure}[t]
  \centering
  \includegraphics[width=1.0\textwidth]{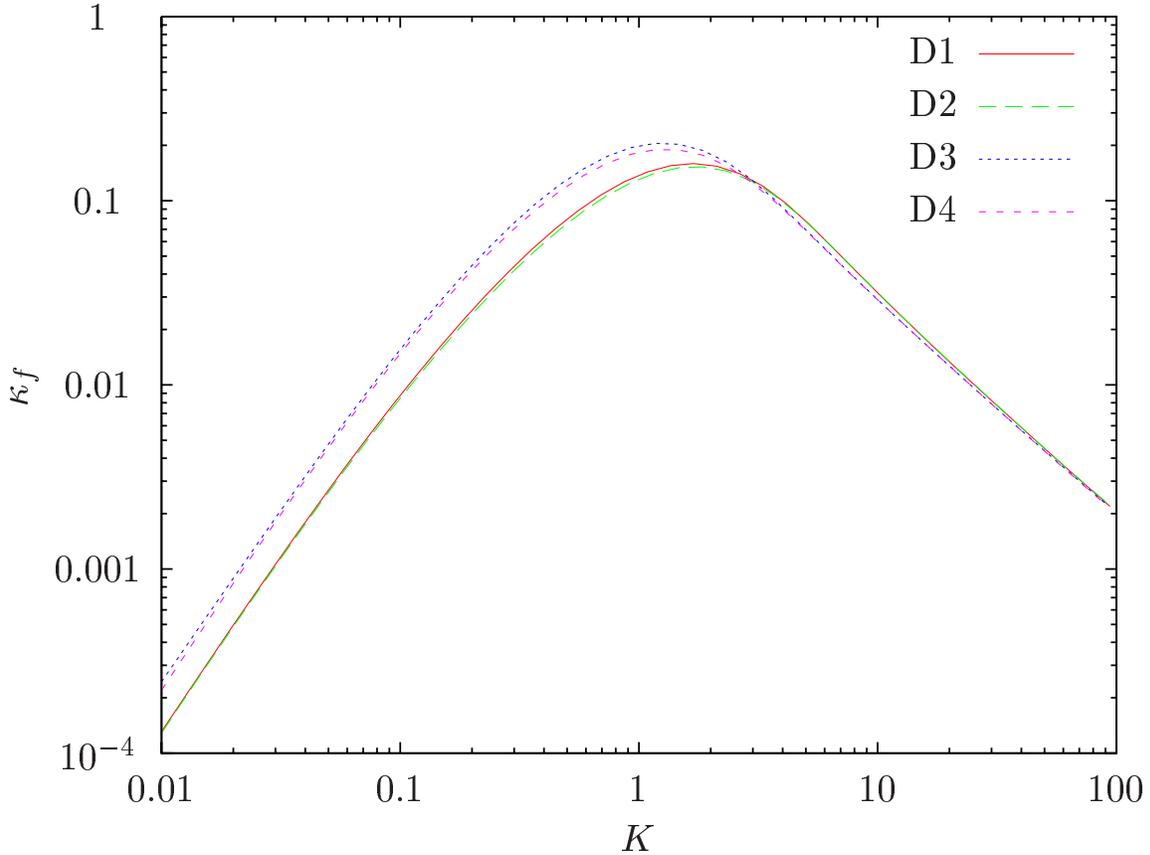} 
  \caption{The final efficiency factor $\kappa$ as a function of $K$
    for the four scenarios within the decay--inverse decay picture,
    assuming a vanishing
    initial RHN abundance.  Solid/red line denotes Case D1, long dashed/green
     D2, dotted/blue D3, and short dashed/magenta D4.
}
  \label{fig:k_fd_mb-full}
\end{figure}

The time evolution of the corresponding absolute value of
the lepton asymmetry is shown in the lower panel of
Figure~\ref{fig:NBL-N_dec}. A negative lepton asymmetry  is
produced at high temperatures by RHN production from inverse
decays. At $z\sim 5$ in the weak wash-out regime ($z\sim 1$ if
strong wash-out), decays come to dominate over inverse decays, thus 
reversing the direction of the asymmetry production, and eventually
flipping
the sign of the asymmetry to positive. When the RHN abundance begins
to fall off exponentially, the asymmetry also asymptotes to a final, 
constant value.

In the weak wash-out regime ($K=0.1$) the asymmetries produced in
Cases D3 and D4 which include quantum statistical factors are always
larger in magnitude than those produced in Cases D1 and D2 which
assume Maxwell--Boltzmann statistics throughout the whole temperature
range considered.  The change of sign also occurs slightly earlier in
D3 and D4.  These effects can be understood as follows.  From, e.g.,
Eq.~(\ref{eq:BE-fa-full-1a}), we see that the production of a negative
lepton asymmetry at high temperatures by inverse RHN decays is
significantly enhanced when we take proper account of the
Bose--Einstein statistics for the Higgs boson.  Like the case of the
RHN abundance, this effect dominates over the phase space suppression
due to the Fermi--Dirac statistics of the lepton and the RHN.  As we
progress to lower temperatures, RHN decays begin to dominate over
inverse decays, thereby reversing the direction of the lepton
asymmetry evolution.  Since quantum statistics speeds up RHN
production and brings its abundance up to the equilibrium threshold
earlier, the transition from decay to inverse decay domination---and
hence the turning point in the asymmetry evolution---also happens
earlier.  As a result, the asymmetry flips sign a little earlier in
Cases D3 and D4 than in D1 and D2, and has more time to grow to a
larger positive value before the exponential fall-off of the RHN
abundance shuts down the asymmetry production.

In the strong wash-out regime ($K=10$), a similar behaviour is also
visible at $z\lesssim 1$. As we progress to lower temperatures,
however, Cases D3 and D4 end up producing less asymmetry than Cases D1
and D2. This is because for $K>1$, the wash-out rate plays a dominant
role in determining the final asymmetry. Here, quantum statistics
enlarges the phase space of the wash-out term from
$f_{\Phi}f_{l-\overline{l}}$ in Eq.~(\ref{eq:BE-fa-ke-1}) to
$(f_{\Phi}+f_{N})f_{l-\overline{l}}$ in Eq.~(\ref{eq:BE-fa-full-1a}),
thus forcing the lepton asymmetry to flip sign even earlier, and
continuing on to dampen it to a slightly smaller positive value.

Again, as with the RHN, the assumption of kinetic equilibrium has
virtually no effect on the asymmetry evolution: the differences
between Cases D1 and D2, and between Cases D3 and D4 are generally at
the percent level, too small to be visible in
Figure~\ref{fig:NBL-N_dec}.

Finally, Figure~\ref{fig:k_fd_mb-full} summarises the lepton asymmetry
produced in the four cases, in terms of the final efficiency factor
$\kappa_{f}$ defined in Eq.~(\ref{eq:ef-fac}), as a function of the
decay parameter $K$.  For all values of $K$ considered, the assumption
of kinetic equilibrium can be seen to produce a minute ($<5\%$)
difference in $\kappa_f$ between Cases D1 and D2 and between Cases D3
and D4.  Quantum statistics, on the other hand, has a generally
stronger effect on the final lepton asymmetry.  In the weak wash-out
regime ($K\lesssim1$), inclusion of quantum statistical factors (Cases
D3 and D4) enhances $\kappa_f$ by a factor of $\sim 1.5$ relative to
Cases D1 and D2 which assume Maxwell--Boltzmann statistics.  In the
strong wash-out regime ($K \gtrsim 1$), the effect of quantum
statistics is to suppress $\kappa_f$ by up to $20$\% at $K \sim 10$,
but reduces to the percent level at $K\sim100$.

\section{Scattering processes}
\label{sec:full-BE-scat}

In this section we enlarge our picture of thermal leptogenesis to
include tree level scattering processes of the RHN with the top quark
, e.g., $Nl \to qt$, which are of
$\mathcal{O}\left(h^{2}_{t}\lambda^{2}\right)$.  These interactions
lead to an additional production channel for the RHN and contribute to
the wash-out processes. Until now these scattering processes have only
been considered using the integrated Boltzmann
equations~\cite{Giudice:2003jh,Buchmuller:2004nz}.  Here we provide
for the first time a solution of the full set of Boltzmann equations
at the mode level.

We do not consider interactions with gauge bosons, nor include
$CP$ violation in $2\rightarrow 2$ or $1(2)\rightarrow 3$ processes,
which are of higher order in the Yukawa couplings.  $CP$ violation
from these processes was considered
in~\cite{Abada:2006ea,Pilaftsis:2003gt,Pilaftsis:2005rv,Nardi:2007jp},
where it was shown that at high temperatures $CP$ violation from
scattering is the main source of lepton asymmetry production.  However
the final asymmetry depends also on the strength of the wash-out
processes; It turns out that in the weak wash-out regime ($K < 1$)
$CP$ violation in the scattering processes tends to suppress the
asymmetry production, while in the transition ($K \simeq 1$) and
strong ($K > 1$) wash-out regimes its contribution is small to
negligible.

In the following we first recall the treatment of scattering processes
in the integrated picture, before we proceed to write down the full
set of mode equations including the relevant scattering terms.
Table~\ref{tab:scat} summarises the assumptions of these two
scenarios.

\begin{table}[t]  
\centering
\caption{Scenarios including scattering with the top quark
  and their associated assumptions.  Case S1 corresponds to the conventional
  integrated Boltzmann approach, while Case S2 involves solving the full
  set of Boltzmann equations at the mode level.}
  \label{tab:scat}
  \begin{tabular}[centered]{|c|c|c|c|c}
    \hline  & Assumption of kinetic equilibrium & Including quantum statistics & Section\\ 
    \hline Case S1 & Yes & No & \ref{subsec:conv-scat}\\
    \hline Case S2 & No & Yes & \ref{subsec:full-scat} \\
    \hline
  \end{tabular}
\end{table}

\subsection{Case S1: Scattering in the integrated picture}
\label{subsec:conv-scat}

Analogous to Eqs.~(\ref{eq:BE-Nn-2}) and (\ref{eq:BE-Na-2}), the
integrated Boltzmann equations including scattering take the following
form~\cite{Buchmuller:2004nz},
\begin{eqnarray}
  \label{eq:BE-Nn-s-1}
  \frac{\partial N_{N}}{\partial z}&=-\,\left(D+S\right)\,
  \left(N_{N}-N_{N}^{\rm eq}\right),\\
  \label{eq:BE-Na-s-1}
  \frac{\partial N_{l-\overline{l}}}{\partial z} &=
 - W\,N_{l-\overline{l}}+\varepsilon\,D\left(N_{N}-N_{N}^{\rm eq}\right),
\end{eqnarray}
where $S$ accounts for the production of RHN from scattering
processes, and the wash-out rate $W$ contains also a contribution from
these processes.  The scattering rate $S$ itself consists of two
terms, $S=2\,S_{s}+4\,S_{t}$, coming respectively from scattering in
the $s$-channel and in the $t$-channel.  One factor of 2 stems from
contribution from processes involving anti-particles, and another
factor of 2 in the $t$-channel term originates from the $u$-channel
diagram.

In general, the scattering rates are defined as
\begin{equation}
  \label{eq:s-int-def}
  S_{s,t}=\frac{\Gamma_{s,t}}{H\,z},
\end{equation}
where
\begin{equation}
  \label{eq:s-int-gamma}
  \Gamma_{s,t}=\frac{M}{24\,\zeta(3)\,g_{N}\,\pi^{2}}\,\frac{\mathcal{I}_{s,t}}{K_{2}(z)\,z^{3}}.
\end{equation}
Note that an additional factor of $4/(3)\,\zeta(3)$ appears
in this definition compared
  to the definition of reference~\cite{Buchmuller:2004nz}. This is due 
to the Fermi--Dirac statistics used in our derivation.
The
quantity $\mathcal{I}_{s,t}$ is an integral
\begin{equation}
  \label{eq:s-int-int}
  \mathcal{I}_{s,t}=\int_{z^{2}}^{\infty}\,d\varPsi\,\hat{\sigma}_{s,t}(\varPsi)\,\sqrt{\varPsi}
  \,K_{1}\left(\sqrt{\varPsi}  \right) 
\end{equation}
of the reduced cross-section $\hat{\sigma}_{s,t}$, given by~\cite{Plumacher:1997ru}
\begin{equation}
  \label{eq:s-int-red-sigma}
  \hat{\sigma}_{s,t}=\frac{3\,h_t^{2}}{4\,\pi}\,\frac{M\,\tilde{m}_{1}}{v^{2}}\,\chi_{s,t}(x),
\end{equation}
where $x=\varPsi/z^{2}$, 
and $h_t= h_t(T)$ is
  the top Yukawa coupling, to be evaluated
at the relevant energy scale (or temperature) $T$
 by solving the
  renormalisation group equation. See Appendix~\ref{sec:RGE}.

The
functions $\chi_{s,t}(x)$ are defined as
\begin{align}
  \label{eq:s-int-fs}
  \chi_{s}(x)&=\left(\frac{x-1}{x}\right)^{2},\\
  \label{eq:s-int-ft}
  \chi_{t}(x)&=\frac{x-1}{x}\,\left[\frac{x-2+2a_{h}}{x-1+a_{h}}+\frac{1-2a_{h}}{x-1}\,
    \log\left(\frac{x-1+a_{h}} {a_{h}}\right) \right],
\end{align}
where we have introduced $a_{h}=m_{\Phi}/M$ as an infrared cut-off for
the $t$-channel diagram, and $m_{\Phi}$ is the mass of the Higgs boson
which presumably receives contributions from interactions with the
thermal bath, i.e., its value does not correspond to that potentially
measured at the LHC.  The value of $m_\Phi$ can in principle be
deduced from a thermal field theoretic treatment of leptogenesis, and
the analysis of~\cite{Giudice:2003jh} found $m_{\Phi}(T)\simeq
0.4\,T$.  However some open questions still remain and hence in the
present work we prefer to adopt a value of $a_{h}=10^{-5}$, used first
by Luty in~\cite{Luty:1992un}.

It is convenient to rewrite the $s$- and $t$-channel scattering rates
$S_{s,t}$ in terms of the functions $f_{s,t}$ defined as
\begin{equation}
  \label{eq:s-int-fst-z}
  f_{s,t}(z)=\frac{\int_{z^{2}}^{\infty}\,d\varPsi\,\chi_{s,t}\left(\varPsi/z^{2}\right)
    \, \sqrt{\varPsi}\,K_{1}\left(\sqrt{\varPsi}\right)}{z^{2}\,K_{2}(z)},
\end{equation}
such that 
\begin{equation}
  \label{eq:s-int-Sst}
  S_{s,t}=\frac{K_{s}}{9\,\zeta(3)}\,f_{s,t},
\end{equation}
and the total scattering rate is given by
\begin{equation}
  \label{eq:s-int-expl}
  S=\frac{2\,K_{s}}{9\,\zeta(3)}\,\left(f_{s}(z)+2\,f_{t}(z)\right),
\end{equation}
where
\begin{equation}
  \label{eq:K-s}
  K_s=\frac{\tilde{m}_{1}}{m_{\ast}^{s}} ,  
\end{equation}
with $\tilde{m}_1$ given by Eqs.~(\ref{eq:m-nu-eff}), and~\cite{Buchmuller:2004nz}
\begin{equation}
  \label{eq:m-mu-eff-s}
  m_{\ast}^{s}=\frac{4\pi^{2}}{9}\frac{g_{N}}{h_t^{2}}\,m_{\ast},
\end{equation}
where $m_\ast$ is defined in in Eq.~(\ref{eq:m-nu-eq}).

Since the scattering processes with the top quark change the lepton
number by one unit, they contribute also to the wash-out of the
asymmetry.  The total wash-out rate is given by
\begin{equation}
W=W_{ID}+W_{\Delta L=1},
\end{equation}
where $W_{ID}$ denotes the contribution from inverse decay defined in
Eq.~(\ref{eq:wid}), and $W_{\Delta L=1}$ from scattering in the $s$-
and $t$- channels,
\begin{equation}
  \label{eq:s-int-wo-1}
  W_{\Delta L=1}=W_{s}+2\,W_{t},
\end{equation}
with 
\begin{equation}
  \label{eq:s-int-wo-s}
  W_{s}=\frac{N_{N}}{N_{N}^{\rm eq}}\,\frac{\Gamma_{s}^{l}}{H\,z}=\frac{N_{N}^{\rm eq}}{N_{l}^{\rm eq}}\,
  \frac{N_{N}}{N_{N}^{\rm eq}}\,S_{s},
\end{equation}
and
\begin{equation}
  \label{eq:s-int-wo-t}
  W_{t}=\frac{\Gamma_{t}^{l}}{H\,z}=\frac{N_{N}^{\rm eq}}{N_{l}^{\rm eq}}\,
  S_{t}.
\end{equation}
 The lepton scattering rates are
  given by $\Gamma_{s,t}^{l}=N_{N}^{\rm eq}/N_{l}^{\rm eq}\,\Gamma_{s,t}$. Using
Eq.~(\ref{eq:wid}), the two contributions $W_{ID}$ and $W_{\Delta
  L=1}$ are related by
\begin{equation}
  \label{eq:int-s-wo-l}
  W_{\Delta L=1}= 2\, W_{ID}\frac{1}{D}\, \left( \frac{
      N_{N}}{N_{N}^{\rm eq}}\, S_{s} +2\, S_{t}\right),
\end{equation}
so that
\begin{equation}
  \label{eq:s-int-wo-tot}
  \begin{aligned}
    W= W_{ID}\,\left[1+\frac{1}{D} \left(
        2\, \frac{N_{N}}{N_{N}^{\rm eq}}\, S_{s} +4\,
        S_{t}\right)\right]
  \end{aligned}
\end{equation}
gives the total wash-out rate.

\begin{figure}[t]
  \centering
  \includegraphics[width=1.0\textwidth]{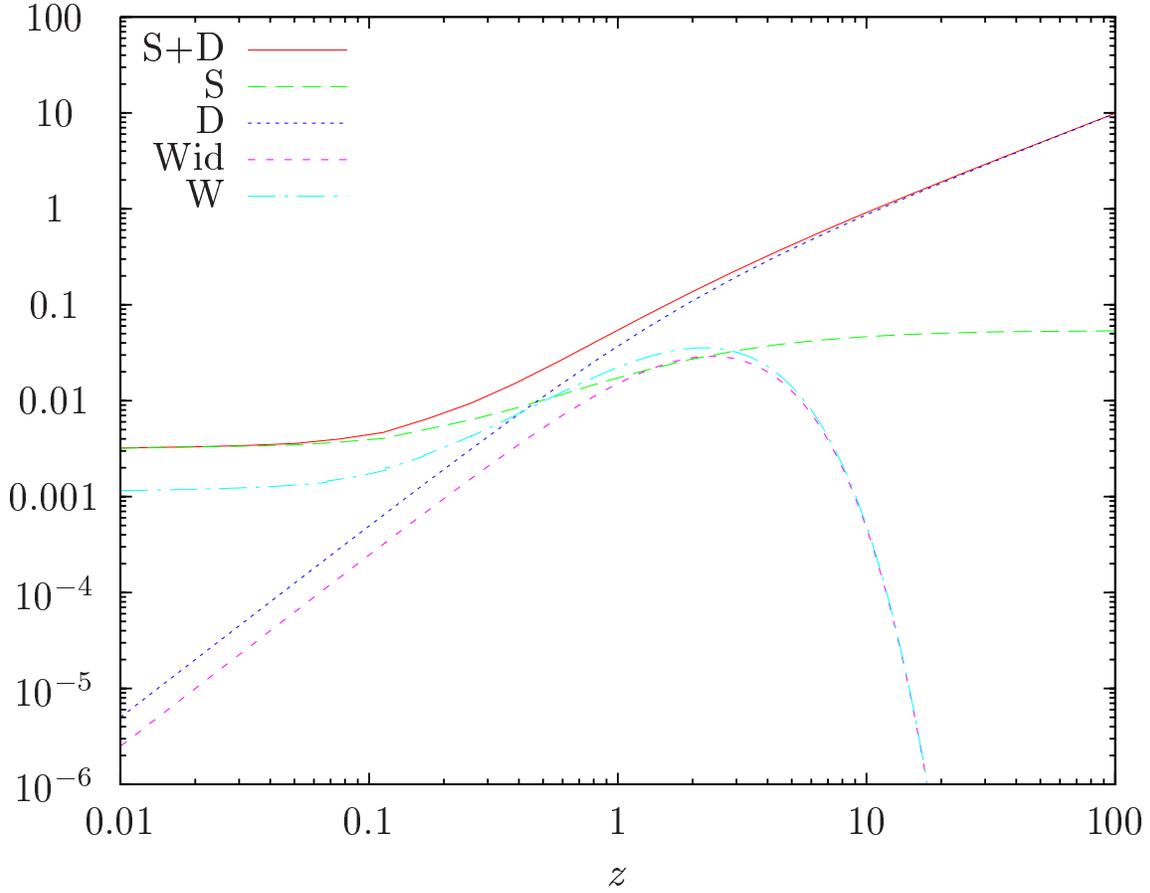} 
  \caption{The decay $D$, scattering $S$, and wash-out rates $W$ and
    $W_{ID}$ as functions of $z$ in the integrated approach, assuming
    $K=0.1$ and $a_{h}=10^{-5}$.}
  \label{fig:dsw}
\end{figure}

Figure~\ref{fig:dsw} shows the various rates $D$, $S$, $W$ and
  $W_{ID}$ as functions of $z$ assuming $K=0.1$. For other choices of
  $K$, the corresponding rates evolve with $z$ in a similar fashion,
  but with magnitudes scaling with $K$.

\subsection{Case S2: Complete mode equations including scattering}
\label{subsec:full-scat}

The basic BE for the distribution function of the RHN is given by
\begin{equation}
  \label{eq:be-rhn-fund-scat}
 \frac{H(M)}{z}\, \frac{\partial f_{N}}{\partial z} 
  =C_{D}\left[f_{N}\right]+2\,C_{S,s}\left[f_{N}\right]+4\,C_{S,t}\left[f_{N}\right],
\end{equation}
where again one factor of 2 stems from contribution from processes
involving anti-particles, and another factor of 2 in the $t$-channel
term originates from the $u$-channel diagram.  The decay--inverse
decay collision integral $C_{D}$ is given in
Eq.~(\ref{eq:be-rhn-fund}), the $s$-channel scattering integral is
\begin{equation}
  \label{eq:CI-full-scat-n-1}
  \begin{aligned}
    C_{S,s}\left[f_{N}\right] &= \frac{1}{2E_{N}} \, \int
    \prod_{i=l,q,t} \frac{dp^{3}_{i}}{(2\pi)^{3}2E_{i}} (2\pi)^{4}
    \delta^{4} (p_{N}+p_{l}- p_{t}-p_{q}) \, \vert \mathcal{M}_{s}
    \vert^{2} \\ &\times \left[
  (1-f_{N})(1-f_{l})f_{t}f_{q} - f_{N}f_{l}(1-f_{t})(1-f_{q}) \right],
  \end{aligned}
\end{equation}
and a similar expression exists for the $t$-channel scattering
integral $C_{S,t}\left[f_{N}\right]$, but with the appropriate matrix
element $\mathcal{M}_{t}$, and the replacements $f_l \leftrightarrow
f_q$.

The analogous equation for the lepton asymmetry is
\begin{equation}
  \label{eq:be-asy-fund-scat}
  \frac{H(M)}{z}\, \frac{\partial f_{l-\overline{l}}}{\partial z} 
  =C_{D}\left[f_{l-\overline{l}}\right]+2\,C_{S,s}\left[f_{l-\overline{l}}\right]
  +4\,C_{S,t}\left[f_{l-\overline{l}}\right],
\end{equation}
where $C_{D}\left[f_{l-\overline{l}}\right] \equiv
C_{D}\left[f_{l}\right]-C_{D}\left[f_{\overline{l}}\right]$ can be
constructed from Eq.~(\ref{eq:be-lep-fund}), and
\begin{equation}
  \label{eq:CI-full-scat-l-1}
  \begin{aligned}
    C_{S,s}\left[f_{l-\overline{l}}\right] &=\frac{1}{2E_{l}}\int
    \prod_{i=N,q,t}
    \frac{dp^{3}_{i}}{(2\pi)^{3}\,2E_{i}}\,(2\pi)^{4}\delta^{4}(p_{l}+p_{N}-p_{q}-p_{t})\;
    \vert \mathcal{M}_{s}\vert^{2} \\ & \times f_{l-\overline{l}}\,
    \left(f_{N}\,\left(f_{t}+f_{q}-1\right)- f_{t}f_{q} \right).
     \end{aligned}
\end{equation}
Replacing $\mathcal{M}_{s}$ with $\mathcal{M}_{t}$ and 
$f_q \leftrightarrow f_N$ in Eq.~(\ref{eq:CI-full-scat-l-1}) 
yields the integral 
$C_{S,t}\left[f_{l-\overline{l}}\right]$.

The collision integrals (\ref{eq:CI-full-scat-n-1}) and
(\ref{eq:CI-full-scat-l-1}) are nine-dimensional and can be reduced
analytically down to two dimensions. We give in
appendix~\ref{sec:reductionCIs} the final reduced integrals and, as an
example, the full reduction procedure applied to the $s$-channel
collisional integral for tracking the RHN following the method
of~\cite{Bolz:2000fu,Pradler:2007ne}.  A general treatment of
scattering kernels in kinetic equations can be found
in~\cite{Hohenegger:2008zj}.

\subsection{Results and discussions}

\subsubsection{Scattering vs decay--inverse decay}

\begin{figure}[t]
  \centering
  \includegraphics[width=1.0\textwidth]{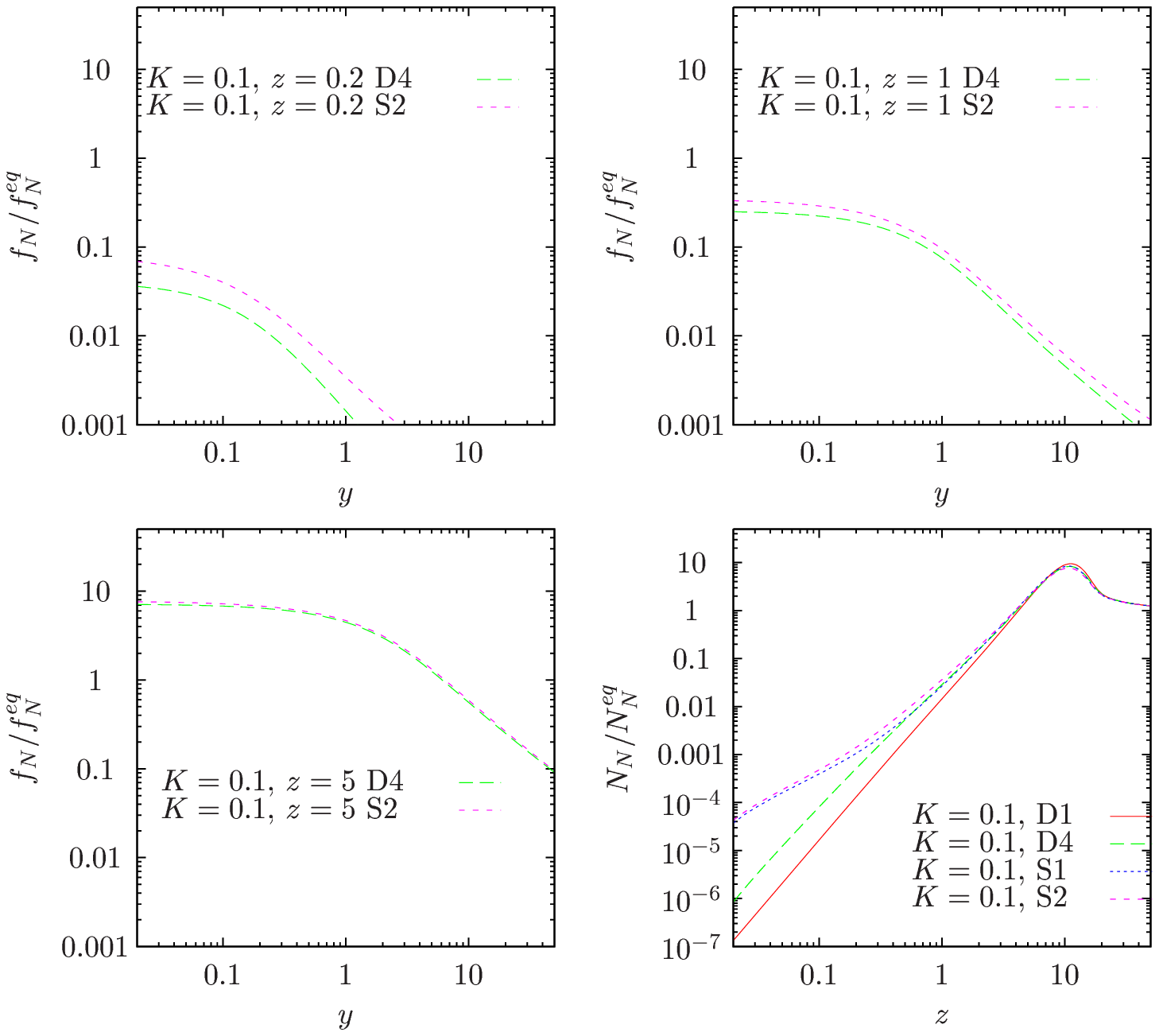} 
  \caption{Snapshots of the RHN distribution function
    $f_{N}/f_{N}^{\rm eq}$ at $z=0.2,1,5$, and the RHN abundance
    $N_{N}/N^{\rm eq}_{N}$ as a function of $z$, assuming $K=0.1$.
    Solid/red line denotes Case D1, long dashed/green D4, dotted/blue
    S1, and short dashed/magenta S2.  See tables~\ref{tab:dec} and
    ~\ref{tab:scat} for a summary of the scenarios.}
  \label{fig:fn-Nn_full-scatt_1}
\end{figure}

\begin{figure}[t]
  \centering
  \includegraphics[width=1.0\textwidth]{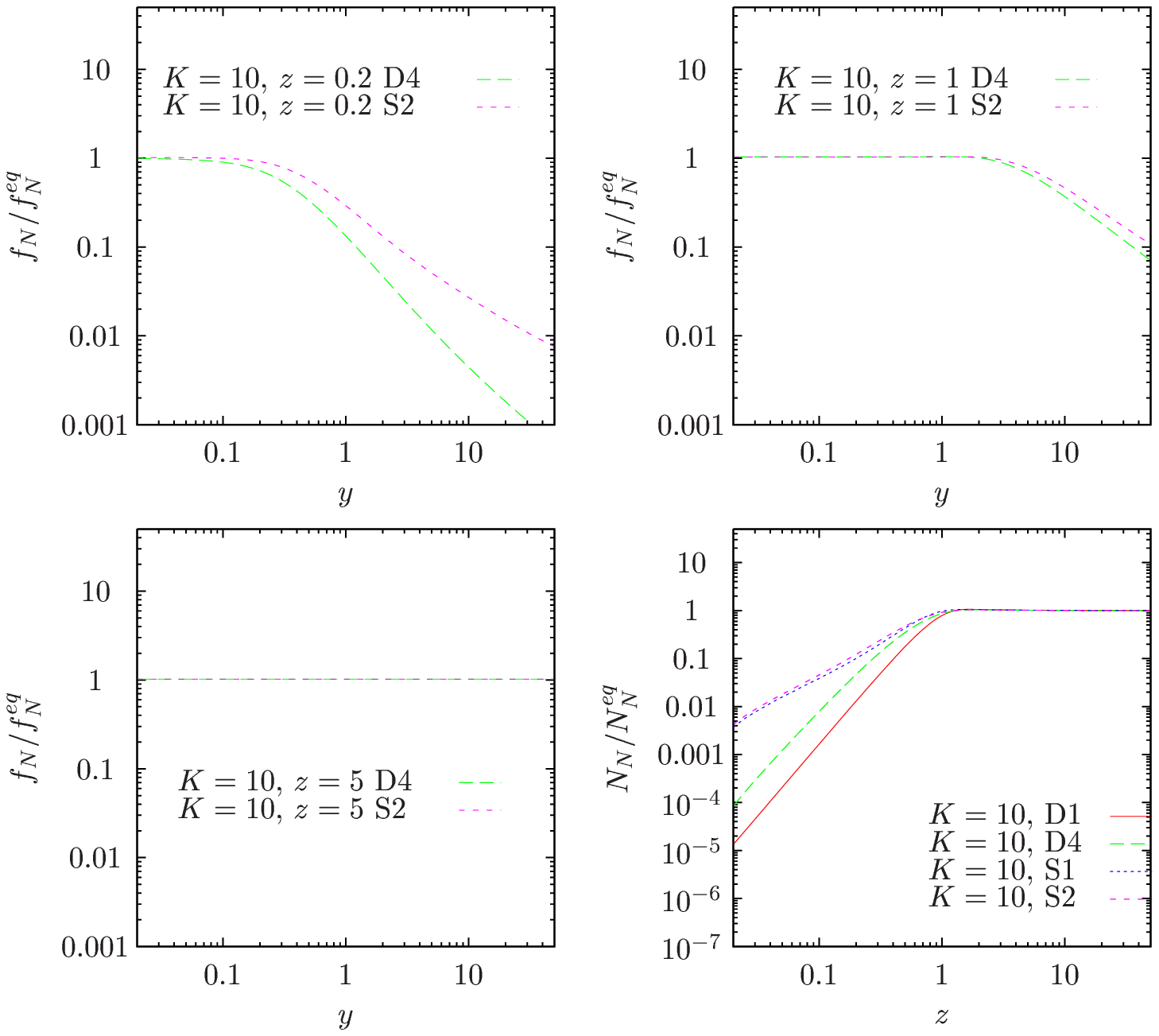}
  \caption{Same as Figure~\ref{fig:fn-Nn_full-scatt_1}, but for
    $K=10$.}
  \label{fig:fn-Nn_full-scatt_2}
\end{figure}

Figure~\ref{fig:fn-Nn_full-scatt_1} shows snapshots of the RHN
distribution function in Case S2 relative to an equilibrium Fermi--Dirac
distribution at time $z=0.2,1,5$, as well as the RHN number density
normalised  to its equilibrium value as a
function of $z$ for an interaction strength of $K=0.1$. These are
compared with their counterparts assuming decay and inverse decay only
(Case D4).  Figure~\ref{fig:fn-Nn_full-scatt_2} is similar, except for
$K=10$.  Clearly, including scattering processes speeds
up the equilibration of the RHN distribution function, especially at
high temperatures ($z<1$).  This effect is more significant for small
values of $K$, since for large $K$ values  decays and
inverse decays are already fast enough to establish equilibrium.

Looking at the time evolution of the RHN number density we see a
corresponding increase in $N_N$ at high temperatures when scattering
is included (Case S2), compared to the case with decays and inverse
decays only (Case D4).  The equilibrium density is also reached at an
earlier time (or higher temperature).  The integrated approach shows a
similar behaviour, with Case S1 predicting a large RHN abundance at
high temperatures and hence faster equilibration than Case D1.

\begin{figure}[t]
  \centering
  \includegraphics[width=1.0\textwidth]{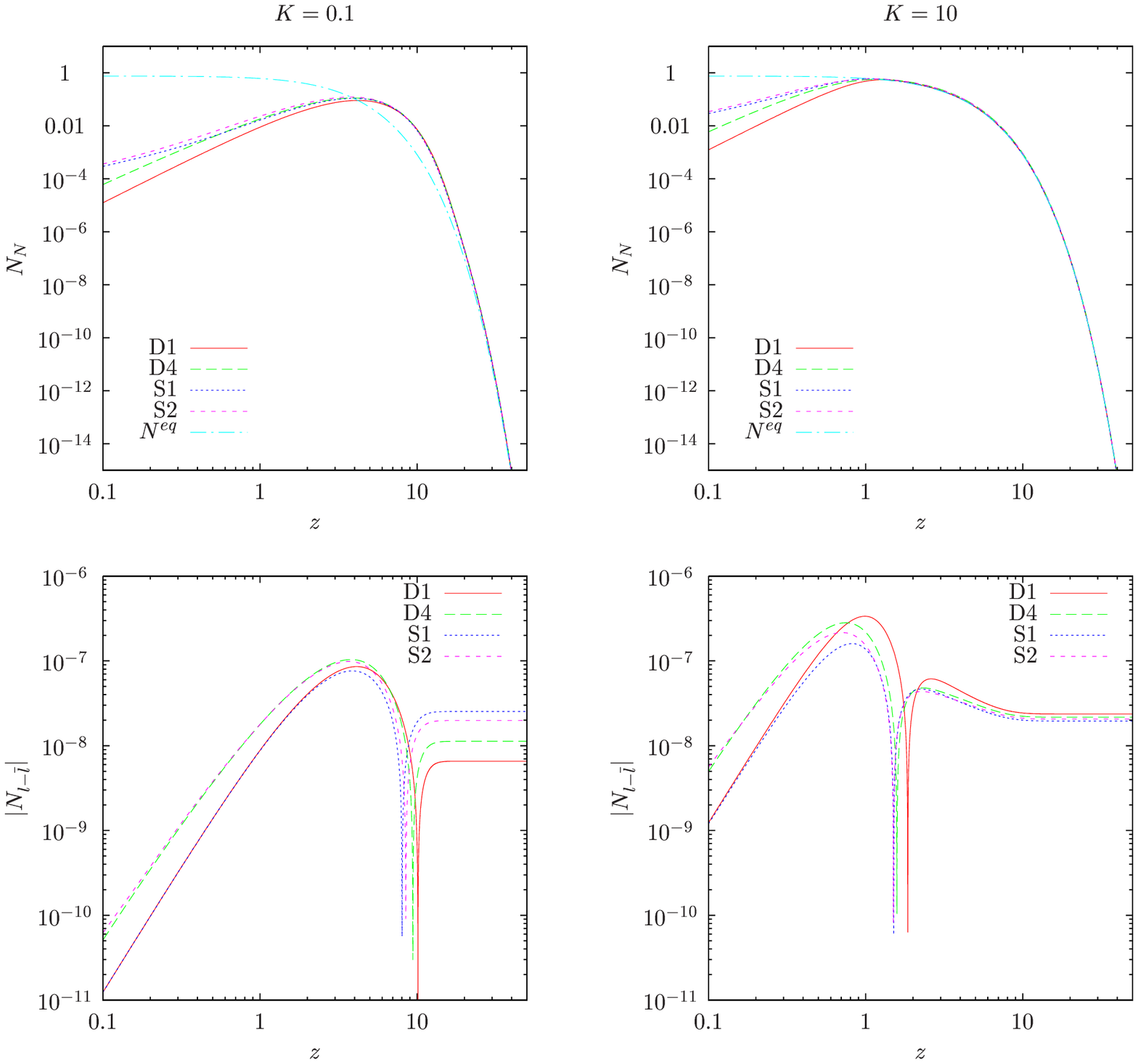}
  \caption{Time evolution of the absolute value of the lepton
    asymmetry $\vert N_{l-\overline{l}} \vert$ for two different
    coupling strengths $K$. Shown are the two cases including
    scattering processes S1 (dotted/blue) and S2 (short
    dashed/magenta), and two scenarios D1 (solid/red) and D4 (long
    dashed/green) within the decays--inverse decay only framework.
    For reference we also plot the RHN equilibrium abundance
    (dot-dash/cyan).}
  \label{fig:NBL_scat}
\end{figure}

Figure~\ref{fig:NBL_scat} shows the time evolution of the lepton
asymmetry, again for the two characteristic values of the decay
parameter $K=0.1,10$. As discussed earlier, we have explicitly ignored
$CP$ violation in the scattering processes, so that they have no
direct influence on the lepton asymmetry.  This assumption is the
reason why, in the weak wash-out regime ($K=0.1$), the asymmetry
evolution at high temperatures ($z<1$) in Case S2 is virtually
identical to that in its decay--inverse decay only counterpart Case
D4.  The same behaviour can also be seen when comparing Cases S1 and
D1.  Here, decays and inverse decays of the RHN alone source the
creation of a lepton asymmetry.  Since for $K<1$ the asymmetry
evolution at high temperatures hinges primarily on inverse decays and
is as yet unaffected by such external factors as the RHN abundance and
wash-out processes, the inclusion of scattering processes has no
visible effect on $N_{l-\overline{l}}$.

However, scattering can still affect the asymmetry production in two
indirect and competing ways: (i) the larger RHN abundance produced via
scattering processes at high temperatures forces the lepton asymmetry
to flip sign earlier, thereby generating a larger positive lepton
asymmetry, and (ii) scattering leads to additional wash-out of the
lepton asymmetry.  The first effect dominates for coupling strengths
lying in the weak wash-out regime ($K<1$), eventually leading to a
larger asymmetry in Cases S1 and S2, compared with their
decay--inverse decay only counterparts D1 and D4 as shown in
Figure~\ref{fig:NBL_scat}.  For stronger couplings ($K>1$), the second
effect dominates; in fact, Figure~\ref{fig:NBL_scat} shows that the
additional wash-out due to scattering suppresses the lepton asymmetry
production in Cases S1 and S2 already at high temperatures $z < 1$,
compared with the decay--inverse decay only scenarios D1
and D4.

\subsubsection{Complete treatment vs integrated approach \label{sec:quantum}}

The complete treatment differs from the integrated approach in that in
the latter case we assume kinetic equilibrium for the RHN and neglect
all quantum statistical factors.  As we saw in section~\ref{sec:results},
the assumption of kinetic equilibrium tends to underestimate by a tiny
amount the RHN abundance at $z<1$.  This can be understood from
Figures~\ref{fig:fn-Nn_full-scatt_1} and \ref{fig:fn-Nn_full-scatt_2}
as a result of the more efficient production of low momentum RHN
states, which in turn contribute more to the momentum integral.

Quantum statistics, on the other hand, has very different effects on
the scattering and the decay--inverse decay collision terms.  As we
saw in section~\ref{sec:results}, in the decay--inverse decay
scenario, quantum statistics always enhances the interaction rates
through the enlarged Higgs boson phase space density $f_\Phi$ at low
$E_\Phi$.  For the scattering processes, since all participants are
fermions, the role of quantum statistics is to reduce the phase space
and hence suppress the interaction rates. In general, however, we
expect quantum statistics to be more important for decay/inverse decay
than for scattering.  This is because in the decay--inverse decay case
the enhanced phase space due to $f_{\Phi}$ at low $E_{\Phi}$ can in
principle be infinite, while Pauli blocking for fermions participating
in scattering, e.g., $1-f_l$, suppresses the phase space by at most a
factor of $1/2$.

The difference between the RHN abundance and the lepton asymmetry
evolution in the complete and the integrated treatments can then be
understood in terms of a competition between the three aforementioned
effects.

Consider first the RHN abundance.  At $z \ll 1$ the dominant RHN
production channel is scattering.  Here, suppression of the production
rate due to quantum statistical factors competes with the small
enhancement due to our dropping the assumption of kinetic equilibrium.
The net result is that both Cases S1 and S2 give very similar RHN
abundances as shown in Figures~\ref{fig:fn-Nn_full-scatt_1} to
\ref{fig:NBL_scat}.  At $z \sim 0.3$, decay/inverse decay becomes
comparable to scattering (see Figure~\ref{fig:dsw}).  Here, the
enhanced decay rate due to quantum statistics in Case S2 pushes up RHN
production relative to Case S1.  This effect is more prominent in the
weak wash-out regime than in the strong wash-out region, since in the
former case the RHN abundance is further away from equilibrium.
Progressing further in $z$, we see that the RHN abundances in Cases S1
and S2 become virtually identical already before reaching the
equilibrium value.  This is in stark contrast with the decay--inverse
decay only scenarios, where the RHN abundances in Cases D1 and D4
clearly cross the equilibrium threshold at different times.

Consider now the evolution of the lepton asymmetry 
(lower panel of Figure~\ref{fig:NBL_scat}).  Comparing
Cases S1 and S2 in the weak wash-out regime ($K=0.1$), quantum
statistics in the latter scenario enhances the production of a
negative lepton asymmetry at high temperatures.  This effect is due
solely to phase space enhancements in the inverse decay term, since we
have assumed explicitly that scattering does not violate $CP$.  At $z
\sim 4$, the production of lepton asymmetry reverses direction as RHN
decays begin to dominate over inverse decays. As mentioned earlier,
quantum statistics causes this reversal to happen earlier in the
decay--inverse decay only scenario by bringing the RHN abundance to
the equilibrium threshold at an earlier time.  When including
scattering, however, the RHN abundances in both Cases S1 and S2 cross
the equilibrium threshold at almost the same time, as discussed in the
previous paragraph.  This means that the evolution of their
corresponding lepton asymmetries also turns around at roughly the same
time.  Since at the time of the turn-around Case S2 has a more
negative asymmetry than Case S1, the net effect is that the asymmetry
in Case S2 flips sign at a slightly later time than in Case S1, and
subsequently grows to a smaller positive value.

The effects of quantum statistics on the lepton asymmetry evolution in
the strong wash-out regime ($K=10$) can be similarly understood,
except that we must consider also the role of the wash-out terms.  At
$z \lesssim 1$, the wash-out rate is dominated by scattering.
However, as shown in Figure~\ref{fig:dsw}, decay/inverse decay becomes
comparable to scattering at $z \sim 0.3$ and is the dominant wash-out
process at $z \gtrsim 1$.  Thus, from $z \sim 1$ onwards, the net
effect of quantum statistics is to enhance the wash-out rate.  This
effect can be seen at the turn-around of the lepton asymmetry
evolution: the stronger wash-out rate in Case S2 forces the lepton
asymmetry evolution to reverse direction at a slightly earlier time
than in Case S1.  However, since at the time of the turn-around Case
S2 has a more negative asymmetry than S1, the asymmetries in both
cases end up flipping signs at almost the same time and grow to nearly
identical values.

\begin{figure}[t]
  \centering
  \includegraphics[width=1.0\textwidth]{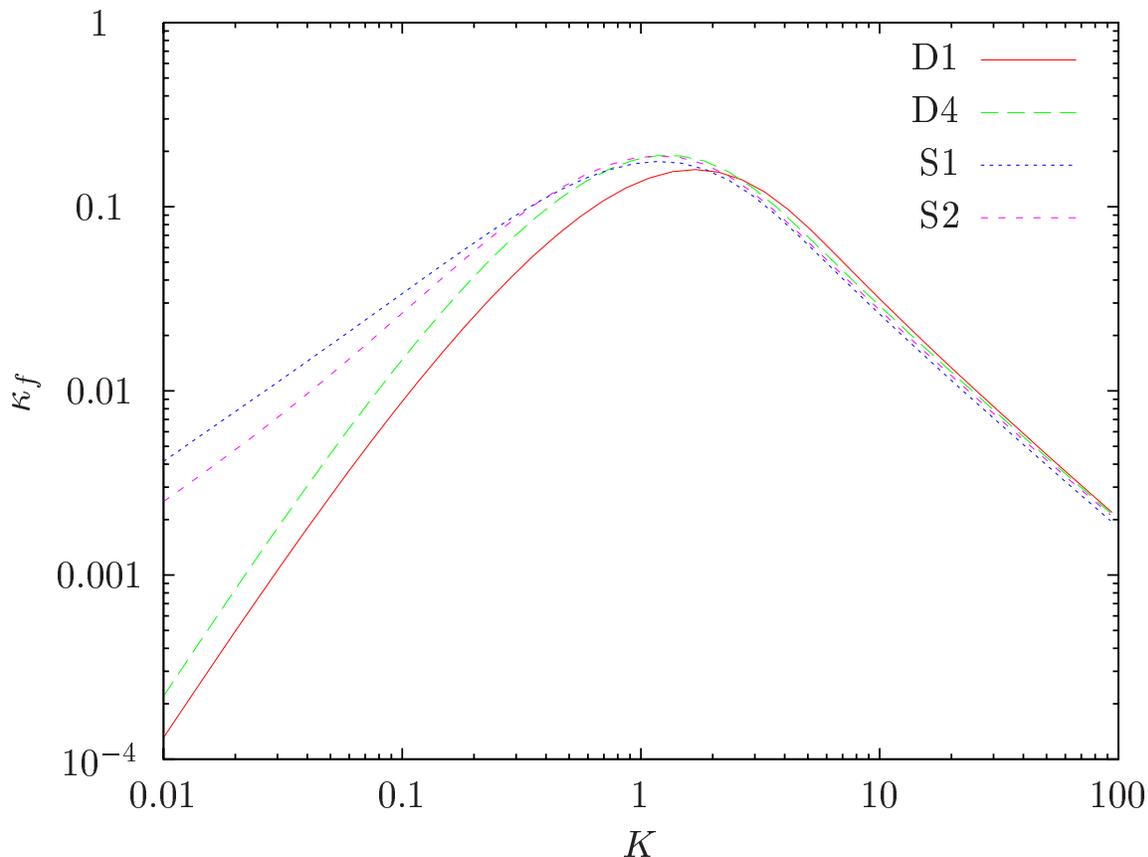}
  \caption{The final efficiency factor $\kappa_f$ without and with
    scattering terms: D1 (solid), D4 (long dashed), S1 (dotted), and
    S2 (short dashed). }
  \label{fig:k_full-scat}
\end{figure}

Finally, Figure~\ref{fig:k_full-scat} shows the final efficiency
factors as a function of $K$ for the integrated approach and the
complete mode treatment, both including and excluding scattering.  For
Cases S1 and S2 which include scattering, we note that their
difference is rather large in the weak wash-out regime ($K<1$), with
the integrated approach overestimating $\kappa_{f}$ by up to a factor
$\sim 1.5$ at $K\sim 0.01$ compared to solving the complete mode
equations. But this difference decreases as we increase $K$.  At
$K \gtrsim3$ the integrated approach
underestimates $\kappa_{f}$ by less than $\sim 10$\%.

It is also interesting to note that the relative contribution of
scattering processes to the final efficiency factor is smaller in the
complete mode calculation than in the integrated approach.  In the
weak wash-out regime, including scattering enhances the final
efficiency factor from decays and inverse decays by up to a factor of
$\sim 30$ in the integrated scenario.  In the complete mode
calculation, however, the enhancement is only a factor of $\sim 15$.
Similarly, in the strong wash-out regime, scattering reduces
$\kappa_f$ by up to $20$\% in the integrated picture, compared
to below $10$\% in the complete treatment.

\section{Conclusions\label{sec:conclusions}}

In this paper we have for the first time studied leptogenesis by means
of the full Boltzmann equations incorporating all quantum statistical terms 
without the assumption of kinetic equilibrium,
and including scatterings of the
right-handed neutrino with quarks. This is of particular relevance for
the creation of the cosmological baryon asymmetry due to the required
deviation from thermal equilibrium and the energy dependence of all
interactions. As the simplest possible set-up to study these effects,
we have considered only an asymmetry being created by the lightest
right-handed neutrino and have neglected potentially important flavour
effects.

In the conventional approach, i.e., neglecting quantum statistical factors
and assuming kinetic equilibrium, considering only decays and inverse
decays is known to give a rather precise approximation of the final
baryon asymmetry in the interesting strong wash-out regime. Hence, we
start by considering this case, which also offers the possibility
to study the influence of various effects separately and in detail.

Interestingly enough, dropping the assumption of kinetic equilibrium has
almost no effect on the evolution of the RHN number density and the
lepton asymmetry. Taking the full energy dependence of interactions 
into account changes the final efficiency factor by 5\% at the most.

Including all quantum statistical factors has somewhat
larger effects. These factors tend to enlarge the phase space available
for neutrino production by inverse decays, thus significantly boosting
the RHN abundance and the ``wrong-sign'' asymmetry being produced at
high temperatures. Further, they lead to an earlier domination
of decays over inverse decays, thus speeding up the production
of the final asymmetry. In the weak wash-out regime this leaves
more time for the production of an asymmetry, thus leading to a boost
in the final asymmetry by $\sim$50\%.

In the strong wash-out regime, on the other hand, the
final asymmetry is suppressed by between 20\% at $K=10$ and 1\% at $K=100$.
This is again due to the enlarged phase space of inverse decay processes
which act as wash-out terms, thus reducing the asymmetry compared
to the case where quantum statistical factors are neglected.

In the case of scatterings of RHNs off quarks, in contrast to 
decays and inverse decays, quantum statistical factors reduce
the phase space available, since all external particles in these
processes are fermions.  Hence, quantum effects generally tend to
reduce the importance of these scatterings.

Nonetheless, at high temperatures ($z<1$) scattering processes
increase the amount of RHNs being produced, thus making leptogenesis
more efficient.  On the other hand, at low temperatures ($z>1$) they
act as wash-out terms thereby reducing the produced asymmetry. The
first effect, i.e., the more efficient production of RHNs dominates in
the weak wash-out regime, thus leading to a larger final lepton
asymmetry compared to the case where only decays and inverse decays
are included. In the strong wash-out regime the effect from increased
wash-out dominates, i.e., including scatterings leads to a somewhat
reduced asymmetry here.

This is qualitatively in line with results obtained in the integrated
picture, i.e., neglecting quantum statistical factors and assuming
kinetic equilibrium. However, since quantum statistical factor enhance
decays and inverse decays while suppressing scatterings of RHNs with
quarks, the net influence of these scattering processes is reduced
when quantum factors are included. This significantly reduces the
spread of results for the final efficiency factor, particularly in the
weak wash-out regime.

\acknowledgments

We thank Steen Hannestad, Josef Pradler, Max Huber and Georg Raffelt
for useful discussions.

\appendix
\section*{Appendices}

\section{Reduction of the scattering collision integrals}
\label{sec:reductionCIs}

\subsection{$s$-channel}
\label{subsec:full-s-channel}

\subsubsection{Right-handed neutrino}\label{sec:right-hand-neutr}

The full collision term for the $s$-channel in
Eq.~(\ref{eq:be-rhn-fund-scat}) is
\begin{equation}
  \label{eq:C-s-n-1}
  C_{S,s}[f_N] = \frac{1}{2E_{N}} \, \int  \prod_{i=l,q,t}
  \frac{d^{3}p_{i}}{(2\pi)^{3}2E_{i}} (2\pi)^{4} \delta^{4} (p_{N}+p_{l}-
  p_{t}-p_{q}) \, \vert \mathcal{M}_{s} \vert^{2}\;
  \Lambda_{s}^{(N)}\left(f_{N},f_{l},f_{t},f_{q} \right) ,
\end{equation}
with phase space factor $\Lambda^{(N)}_s$ given by
\begin{equation}
  \label{eq:lam-s-n-1}
  \Lambda_{s}^{(N)}\left(f_{N},f_{l},f_{t},f_{q} \right)=\left[
    (1-f_{N})(1-f_{l})f_{t}f_{q} - f_{N}f_{l}(1-f_{t})(1-f_{q})
  \right].
\end{equation}
The matrix element $\mathcal{M}_{s}$ is summed over all internal
degrees of freedom of the particles in the initial and final states,
including colour and isospin, and is given by
\begin{equation}
  \label{eq:C-s-n-3}
  \vert \mathcal{M}_{s} \vert ^{2}= 24\,h_t^{2}
  \frac{M\tilde{m}_{1}}{v^{2}} \, \frac{p_{N}p_{l} 
    \; p_{t}p_{q}}{s^{2}},
\end{equation}
where $\tilde{m}_{1}=\left(m^{\dagger}_{D}m_{D}\right)/M$ is the
effective neutrino mass~\cite{Plumacher:1996kc}, $v=174\;\GeV$ the
vacuum expectation value of the Higgs field and $h_{t}^{2}$ the
top Yukawa coupling given in Appendix~\ref{sec:RGE}.

We work in the centre-of-mass frame, i.e.,
\begin{equation}
  \label{eq:C-s-n-4}
   \mathbf{p}_{N}+\mathbf{p}_{l}=\mathbf{p}_{t}+\mathbf{p}_{q} \equiv  \mathbf{q}.
\end{equation}
In general, the 4-vector delta function can dealt with using the relation
\begin{equation}
  \label{eq:C-s-n-5}
  \begin{aligned}
  \delta \left(p_{i}^{2}-M_{i}^{2}\right) &=\delta \left( E_{i}^{2}-\left(
  \vert \mathbf{p}_{i} \vert^{2} +M_{i}^{2} \right) \right) \\ &= \left (
    \frac{\delta \left (E_{i}
        -\sqrt{\vert \mathbf{p}_{i}\vert^{2}+M_{i}^{2}}
      \right)}{2\sqrt{\vert \mathbf{p}_{i} \vert^{2}+M_{i}^{2}}} + \frac{\delta \left (E_{i}
        + \sqrt{\vert \mathbf{p}_{i} \vert^{2}+M_{i}^{2}}
      \right)}{2\sqrt{\vert \mathbf{p}_{i} \vert^{2}+M_{i}^{2}}} \right ).
  \end{aligned}
\end{equation}
Using this relation, and the fact that 
we consider all particles except the RHN to be massless, i.e., 
\begin{equation}
  \label{eq:C-s-n-6}
\begin{aligned}
  E_{l,t,q}&=\vert \mathbf{p}_{l,t,q} \vert, \\
   E_{N}&=\sqrt{\vert \mathbf{p}_N\vert^{2}+M^{2}},
\end{aligned}
\end{equation}
we can integrate over the quark energy,
\begin{equation}
  \label{eq:C-s-n-7}
  \begin{aligned}
    \int \frac{d^{3} p_{q}}{2E_{q}}\; \delta^{4}
    (p_{N}+p_{l}-p_{t}-p_{q}) &=\int dE_{q}\,
    d^{3}\,p_{q}\,\frac{\delta \left (E_{q}-\vert \mathbf{p}_{q} \vert
      \right )}{2 \vert \mathbf{p}_{q} \vert } \; \Theta (E_{q})\,
    \delta ( E_{N}+E_{l}-E_{t}-E_{q} ) \\ & \hphantom{\int
      d^{3}}\times \delta^{3} \left(
      \mathbf{p}_{N}+\mathbf{p}_{l}-\mathbf{p}_{t}-\mathbf{p}_{q}
    \right ) \\ &= \frac{\delta \left ( E_{N}+E_{l}-E_{t}- \vert
        \mathbf{p}_{N}+\mathbf{p}_{l}-\mathbf{p}_{t} \vert \right ) }
    {2\, \vert \mathbf{p}_{N}+\mathbf{p}_{l}-\mathbf{p}_{t} \vert }\;
    \Theta ( E_{N}+E_{l}-E_{t}) \\ &= \frac{\delta \left (
        E_{N}+E_{l}-E_{t} - \vert \mathbf{q}-\mathbf{p}_{t} \vert
      \right ) }{2\, \vert \mathbf{q} -\mathbf{p}_{t} \vert }\; \Theta
    ( E_{N}+E_{l}-E_{t}) \\ &= \delta \left ( (E_{N}+E_{l}-E_{t})^{2}
      - \vert \mathbf{q}-\mathbf{p}_{t} \vert^{2} \right )\; \Theta (
    E_{N}+E_{l}-E_{t}).
  \end{aligned}
\end{equation}
Similarly, we can rewrite 
\begin{equation}
  \label{eq:C-s-n-8}
  \begin{aligned}
    \frac{d^{3} p_{l}}{2E_{l}} 
&= \int dE_l \, \frac{ \delta \left( E_{l}-\vert
        \mathbf{p}_{l}\vert \right)}{2 \vert
      \mathbf{p}_{l} \vert } 
    \Theta (E_{l})\,  d^{3}p_{l} \\
&= \int dE_l \, \delta \left(
      E^{2}_{l}- \vert \mathbf{p}_{l}
      \vert^{2} \right)\, \Theta (E_{l})\,d^{3}p_l \\
 &= \int dE_l \, \delta \left(
      E^{2}_{l}- \vert \mathbf{q}-\mathbf{p}_{N}
      \vert^{2} \right)\, \Theta (E_{l})\,d^{3}q, \\
  \end{aligned}
\end{equation}
where the last equality follows from changing the variable
from $\mathbf{p}_l$ to  $\mathbf{q}=\mathbf{p}_N+\mathbf{p}_l$, 
and hence $d^3p_l$ to $d^3q$,
and the integration is over the lepton energy $E_l$.

We choose an explicit coordinate system,
\begin{equation}
  \label{eq:C-s-n-9}
  \begin{aligned}
   \mathbf{q} &= \vert \mathbf{q} \vert\,(0,0,1), \\
   \mathbf{p}_{N} &= \vert \mathbf{p}_{N}
   \vert \, (0, \sin \eta , \cos \eta  ), \\
   \mathbf{p}_{t} &= E_{t} (\cos \phi\, \sin \vartheta, \sin \phi \,
   \sin \vartheta , \cos \vartheta),
  \end{aligned}
\end{equation}
and obtain the following quantities: 
\begin{equation}
  \label{eq:C-s-n-10}
  \begin{aligned}
    s &= \left( p_{N}+p_{l} \right )^{2} = \left( p_{t}+p_{q}
    \right)^{2}= \left ( E_{N}+E_{l} \right)^{2} -\vert
    \mathbf{q} \vert^{2}, \\ 
p_{N}p_{l}  &=\frac{s-M^2}{2}, \\ 
p_{q}p_{t}  &= \frac{s}{2}, \\
    \vert
    \mathbf{q}-\mathbf{p}_{t} \vert^{2} &= \vert
    \mathbf{q} \vert^{2}+\vert \mathbf{p}_{t}
    \vert^{2}-2\mathbf{q} \cdot \mathbf{p}_{t} =
    \vert \mathbf{q}\vert^{2}+E^{2}_{t} -2\vert
    \mathbf{q} \vert\,E_{t}\,\cos \vartheta,\\
    \vert \mathbf{q}-\mathbf{p}_{N} \vert^{2}&=
    \vert \mathbf{q} \vert^{2} +
    \vert \mathbf{p}_{N}\vert^{2}-2\mathbf{q} \cdot \mathbf{p}_{N}
=\vert \mathbf{q} \vert^{2}+\vert \mathbf{p}_{N}\vert^{2}-2\,\vert
    \mathbf{q} \vert \, \vert \mathbf{p}_{N} \vert
    \,\cos \eta.
  \end{aligned}
\end{equation}
The matrix element in these coordinates reads
\begin{equation}
  \label{eq:s-n-matrix-3}
  \vert \mathcal{M}_{s} \vert ^{2}=  6\,h_t^{2}
  \frac{M\tilde{m}_{1}}{v^{2}} \,
  \frac{\left(E_{N}+E_{l}\right)^{2}-M^{2}-\vert\mathbf{q}\vert^{2}
  }{\left(E_{N}+E_{l}\right)^{2}
      -\vert \mathbf{q} \vert^{2}}, 
\end{equation}
and the delta functions are given by 
\begin{equation}
  \label{eq:C-s-n-11}
  \begin{aligned}
    \delta \left ( \left( E_{N}+E_{l}-E_{t} \right)^{2} -\vert
      \mathbf{q}-\mathbf{p}_{t} \vert^{2} \right) &= \delta \left(
      \left( E_{N}+E_{l}-E_{t} \right)^{2} -\vert \mathbf{q} \vert^{2}
      -E^{2}_{t} +2\vert \mathbf{q} \vert\, E_{t}\,\cos \vartheta
    \right) \\ &= \frac{1}{2\vert \mathbf{q} \vert \,E_{t}}\; \delta
    \left( \cos \vartheta - \frac{E^{2}_{t}-\left( E_{N}+E_{l}-E_{t}
        \right)^{2} +\vert \mathbf{q} \vert^{2} }{2\vert
        \mathbf{q}\vert \,E_{t}} \right ), \\ \delta \left( E^{2}_{l}
      -\vert \mathbf{q}-\mathbf{p}_{N} \vert^{2} \right) &= \delta
    \left( E^{2}_{l}-\vert \mathbf{q} \vert^{2}-\vert \mathbf{p}_{N}
      \vert^{2} +2\vert
      \mathbf{q}\vert \vert \mathbf{p}_{N} \vert\,\cos \eta \right) \\
    &= \frac{1}{2\vert \mathbf{q}\vert \vert \mathbf{p}_{N} \vert } \;
    \delta \left( \cos \eta - \frac{\vert \mathbf{p}_{N} 
        \vert^{2}-E_l^2 + \vert \mathbf{q} \vert^{2}}{2\vert
        \mathbf{q}\vert \, \vert \mathbf{p}_{N} \vert}\right).
  \end{aligned}
\end{equation}
Collecting all terms we get for the collision integral 
\begin{equation}
  \label{eq:C-s-n-bl}
  \begin{aligned}
    C_{S,s} [f_N]=& \frac{1}{2\,E_{N}\,(2\pi)^{5}} \int
    \,\frac{d\Omega_{N}}{4\pi}\,d\cos \vartheta\,d\phi \,
    \frac{E^{2}_{t}}{2E_{t}}\, dE_{t}\,dE_{l}\, d^{3}q\,
    \frac{1}{2\vert \mathbf{q} \vert \, \vert
      \mathbf{p}_{N}\vert}\;  \frac{1}{2\vert
      \mathbf{q} \vert\,E_{t}} \; \vert \mathcal{M}_{s} \vert^{2}\\
    &
    \times \delta \left( \cos \eta-\frac{\vert
        \mathbf{p}_{N}\vert^{2} - E_l^2 +\vert \mathbf{q} \vert^{2}}{2\vert
        \mathbf{q} \vert\, \vert \mathbf{p}_{N} \vert} \right)
        \delta
    \left(\cos \vartheta-\frac{E^{2}_{t}-(E_{l}+E_{N}-E_{t})^{2}+\vert
        \mathbf{q} \vert^{2}}{2\vert \mathbf{q} \vert\, E_{t}} \right)
    \\ &
    \times \Lambda_{s}^{(N)} \left(f_{N},f_{l},f_{t},f_{q}\right)\,
   \Theta(E_{l})\, \Theta(E_{t})\,
\Theta(E_{l}+E_{N}-E_{t}) \\ =&
    \frac{1}{2^{7}\pi^{3}\,E_{N}\,\vert \mathbf{p}_{N}\vert} \int
    d\cos \vartheta \, d\cos \eta \, dE_{t}\,
    dE_{l}\,
    d\vert \mathbf{q}\vert   \; \vert \mathcal{M}_{s} \vert^{2}\\
     &\times \delta \left(
      \cos \eta-\frac{\vert \mathbf{p}_N \vert^2-E_l^2+\vert \mathbf{q}
        \vert^{2}}{2\vert \mathbf{q}
        \vert\,\vert \mathbf{p}_N \vert} \right) 
   \delta \left(
      \cos \vartheta- \frac{E_{t}^2-\left(E_{l}+E_{N}-E_{t} 
        \right)^{2}+\vert \mathbf{q}\vert^2}{2\vert
        \mathbf{q} \vert\, E_{t}} \right) \\
    &\times \Lambda_{s}^{(N)} \left( f_{N},f_{l},f_{t},f_{q} \right)\,
    \Theta(E_{l})\, \Theta(E_{t})\,
\Theta(E_{l}+E_{N}-E_{t}).
  \end{aligned}
\end{equation}
Here, in the first equality, we take $d^3p_t = E^2_t d E_t \, d \cos
\vartheta \, d\phi$, and average over the direction of the incoming
RHN by integrating over $d\Omega_{N}/(4\pi)$, where
$d\Omega_{N}=d\theta\,d\cos\eta$, because of rotational invariance
(cf.\ Eq~(\ref{eq:C-s-n-9})). In the
second equality, we take $d^{3}q=4\pi\,\vert \mathbf{q}\vert^2
d\vert\mathbf{q}\vert$, and integrate over all azimuthal angles.

The two remaining angles $\vartheta$ and  $\eta$ run in the range
\begin{equation}
  \label{eq:theta-s-n-1}
  \cos \vartheta, \cos \eta \in [-1,1].
\end{equation}
Since apart from the delta functions the integrand does not depend on
either angle, integrating over these ranges effectively lead to 
new integration limits for the $q$-integral: 
\begin{equation}
  \label{eq:theta-s-n-2}
  \begin{aligned}
    \cos \vartheta&= 1 \quad \Rightarrow \quad
    q \in [E_{l}+E_{N},-E_{l}-E_{N}+2E_{t}], \\
    \cos \vartheta&=-1 \quad \Rightarrow \quad
    q\in [-E_{l}-E_{N},E_{l}+E_{N}-2E_{t}],\\
    \cos \eta&= 1 \quad \Rightarrow \quad
    q\in[-E_{l}+p_{N},E_{l}+p_{N}], \\
    \cos \eta&=-1 \quad \Rightarrow \quad q \in[-E_{l}-p_{N},E_{l}-p_{N}],
  \end{aligned}
\end{equation}
where $p_N \equiv \vert \mathbf{p}_N \vert$, and $q \equiv \vert \mathbf{q} \vert$.
Putting these conditions together we get
\begin{equation}
  \label{eq:theta-s-n-6}
  \sup \left[ \vert 2E_{t}-E_{l}-E_{N} \vert,  \vert
    E_{l}-p_{N} \vert \right] \leq q \leq \inf
  \left[E_{l}+E_{N},E_{l}+p_{N} \right].
\end{equation}
Since $E_{l}+E_{N}>E_{l}+p_{N}$ this reduces to
\begin{equation}
  \label{eq:theta-s-n-7}
  \sup \left[ \vert 2E_{t}-E_{l}-E_{N} \vert,  \vert
    E_{l}-p_{N} \vert \right] \leq q \leq E_{l}+p_{N}.
\end{equation}
Thus, the integration over $\cos \eta$ and $\cos \vartheta$
effectively gives rise to a combination of $\Theta$ functions in the
remaining $3$-dimensional integral.  Together with existing $\Theta$
functions in Eq.~(\ref{eq:C-s-n-bl}), we define
\begin{equation}
  \label{eq:theta-s-n-8}
    \Omega \equiv   \Theta \left(q-\vert 2E_{t}-E_{l}-E_{N}\vert \right)\; \Theta
    \left(q- \vert E_{l}-p_{N} \vert \right)\;
    \Theta \left(E_{l}+p_{N}-q \right) 
    \Theta(E_{l}+E_{N}-E_{t})\; 
\end{equation}
to collectively denote all $\Theta$ functions appearing in the
remaining integral.  Note that we have omitted writing out
$\Theta(E_l) \Theta(E_t)$, since positive particle energies are
understood.

Next, we use the relations
\begin{equation}
  \label{eq:theta-s-n-9}
  \Theta \left(q- \vert 2E_{t}-E_{l}-E_{N} \vert \right)= 1- \Theta
  \left( \vert 2E_{t}-E_{l}-E_{N} \vert-q \right),
\end{equation}
and 
\begin{equation}
  \label{eq:theta-s-n-10}
  \Theta \left(\vert 2E_{t}-E_{l}-E_{N} \vert-q \right)\; \Theta
  \left(E_{l}+p_{N}-q\right) = \Theta \left(\vert 2E_{t}-E_{l}-E_{N} 
    \vert-q \right),
\end{equation}
the latter following from the fact that $\Omega$ in
Eq.~(\ref{eq:theta-s-n-8}) vanishes unless $\vert 2E_{t}-E_{l}-E_{N}
\vert < E_{l}+p_{N}$.  With these we split the function $\Omega$ into
two parts (i.e., $\Omega=\Omega_1+\Omega_2$):
\begin{align}
  \label{eq:theta-s-n-11}
  \Omega_{1} &= \Theta (E_{l}+E_{N}-E_{t}) \; 
\Theta \left( q-\vert E_{l}-p_{N}\vert \right)\;
  \Theta \left( E_{l}+p_{N}-q \right), \\ \label{eq:theta-s-n-12}
  \Omega_{2} &=- \; \Theta (E_{l}+E_{N}-E_{t}) \; 
\Theta \left( q-\vert E_{l}-p_{N}\vert \right)\;
  \Theta \left( \vert 2E_{t}-E_{l}-E_{N} \vert -q \right).
\end{align}
Equation~(\ref{eq:theta-s-n-11}) can be further split into two parts
at $\Theta \left( q- \vert E_{l}-p_{N} \vert \right)$ using the
relation
\begin{equation}
  \label{eq:theta-s-n-13}
   \Theta \left(E_{l}-p_{N} \right) + \Theta
  \left(p_{N}-E_{l} \right)=1,
\end{equation}
from which we find
\begin{align}
  \label{eq:theta-s-n-14}
  \Omega_{1a}&=\Theta (E_{l}+E_{N}-E_{t}) \; \Theta\left( q-
    (E_{l}-p_{N})\right)\; \Theta \left( E_{l}+p_{N}-q \right)\;
  \Theta \left(E_{l}-p_{N} \right),\\
  \label{eq:theta-s-n-15}
  \Omega_{1b}&=\Theta (E_{l}+E_{N}-E_{t}) \; \Theta\left( q-
    (p_{N}-E_{l})\right)\; \Theta \left( E_{l}+p_{N}-q \right)\;\Theta
  \left(p_{N}-E_{l} \right),
\end{align}
so that  $\Omega_1 = \Omega_{1a}+\Omega_{1b}$.

Similarly, Eq.~(\ref{eq:theta-s-n-12}) can be split at $\Theta \left(
  q- \vert E_{l}-p_{N} \vert \right)$ into two parts, $\Omega_2 =
\Omega_{2a} + \Omega_{2b}$, by way of the
relation~(\ref{eq:theta-s-n-13}):
\begin{align}
  \label{eq:theta-s-n-16}
  \Omega_{2a}&=-\; \Theta (E_{l}+E_{N}-E_{t})\; \Theta \left(q-
    (E_{l}-p_{N}) \right) \Theta \left(\vert 2E_{t}-E_{l}-E_{N} \vert
    -q\right)\; \Theta \left(
    E_{l}-p_{N} \right)
  ,\\
  \label{eq:theta-s-n-17}
  \Omega_{2b}&=-\; \Theta (E_{l}+E_{N}-E_{t})\; \Theta\left( q-
    (p_{N}-E_{l})\right)\; \Theta \left(\vert 2E_{t}-E_{l}-E_{N} \vert
    -q\right)\;\Theta \left(p_{N} -E_{l} \right).
\end{align}
One further split is possible at  $\Theta \left( \vert 2E_{t}-E_{l}-E_{N} \vert-q \right)$ using
\begin{equation}
  \label{eq:theta-s-n-18}
 \Theta\left(E_l+E_N - 2E_t\right)+ \Theta \left(2E_{t}-E_{l}-E_{N} \right)=1.
\end{equation}
Putting this relation in Eqs.~(\ref{eq:theta-s-n-16}) and
(\ref{eq:theta-s-n-17}), we find the combinations
\begin{equation}
  \label{eq:theta-s-n-19}
  \begin{aligned}
    \Theta \left(q-(E_{l}-p_{N}) \right) \; \Theta
    (E_{l}+E_{N}-2E_{t}-q) &\Rightarrow E_{t}\leq \frac{1}{2}\left(
      E_{N}+p_{N} \right), \\ 
\Theta
    \left(q-(E_{l}-p_{N}) \right) \; \Theta
    (2E_{t}-E_{l}-E_{N}-q) &\Rightarrow  E_t \geq \frac{1}{2}\left( 2E_{l}+
      \left( E_{N}-p_{N}\right) \right),\\
\Theta \left(q-(p_{N}-E_l) \right) \;
    \Theta (E_{l}+E_{N}-2E_{t}-q) &\Rightarrow E_{t}\leq
    \frac{1}{2}\left( 2E_{l}+ \left(
        E_{N}-p_{N}\right)
    \right),\\ 
\Theta \left(q-(p_{N}-E_l) \right) \; \Theta
    (2E_{t}-E_{N}-E_{l}-q) &\Rightarrow E_t \geq \frac{1}{2}\left(
      E_{N}+p_{N} \right),
  \end{aligned}
\end{equation}
with which we can write down the four parts of $\Omega_2$:
\begin{align}
  \label{eq:theta-s-n-20}
  \Omega_{2a,i}&=- \; \Theta \left(\frac{1}{2} (E_{N}+p_{N})-E_{t}
  \right)\; \Theta \left(q- (E_{l}-p_{N}) \right)\; \Theta
  \left(E_{l}+E_{N}- 2E_{t} -q\right)\nonumber \\ &  \hphantom{=-\;\;}
 \times \Theta (
  E_{l}-p_{N}),\\
  \label{eq:theta-s-n-21}
  \Omega_{2a,ii}&=-\; \Theta (E_{l}+E_{N}-E_{t})\; \Theta \left(E_{t}-
    \frac{1}{2} (2E_{l}+E_{N}-p_{N}) \right)\;\Theta
  \left(2E_{t}-E_{l}-E_{N} -q\right) \nonumber \\ &\hphantom{=-\;\;}
  \times \Theta \left(q- (E_{l}-p_{N}) \right)\; \Theta (
  E_{l}-p_{N}),\\
  \label{eq:theta-s-n-22}
  \Omega_{2b,i}&=-\; \Theta \left( \frac{1}{2}
    (2E_{l}+E_{N}-p_{N})-E_{t} \right) \Theta \left(E_{l}+E_{N}-
    2E_{t} -q\right)\nonumber \\&\hphantom{=-\;\;} \times \Theta\left( q-
    (p_{N}-E_{l})\right)\; \Theta (p_{N} -E_{l})\
  ,\\
  \label{eq:theta-s-n-23}
  \Omega_{2b,ii}&=-\; \Theta (E_{l}+E_{N}-E_{t})\; \Theta
  \left(E_{t}-\frac{1}{2} (E_{N}+p_{N}) \right)\;\Theta
  \left(2E_{t}-E_{l}-E_{N} -q\right)  \nonumber \\& \hphantom{=-\;\;}
  \times \Theta \left(q- (p_{N}-E_{l}) \right)\; \Theta (p_{N} -E_{l}),
\end{align}
such that  $\Omega_{2a}=\Omega_{2a,i}+\Omega_{2a,ii}$
and $\Omega_{2b}= \Omega_{2b,i} + \Omega_{2b,ii}$.

Finally, collecting all terms we obtain the relation
\begin{equation}
  \label{eq:theta-s-n-24}
  \Omega=\sum_\mu \Omega_\mu = \Omega_{1a}+\Omega_{1b}+\Omega_{2a,i}
  +\Omega_{2a,ii}+\Omega_{2b,i}+\Omega_{2b,ii},
\end{equation}
so that the remaining $3$-dimensional integration in the collision
integral~(\ref{eq:C-s-n-bl}) can be equivalently written as
\begin{align}
  \label{eq:C-s-n-13}
  C_{S,s}[f_N] &= \sum_\mu \frac{1}{2^{7}\pi^{3}\,E_{N}\,\vert \mathbf{p}_{N} \vert}
  \int dE_{t}\, dE_{l}\, dq \, \vert \mathcal{M}_{s}
  \vert^{2}\;\Lambda_{s}^{(N)}\left(f_{N},f_{l},f_{t},f_{q}\right)\,
  \Omega_\mu.
\end{align}
The phase space factor reads
\begin{equation}
  \label{eq:lam-s-n-2}
  \Lambda_{s}^{(N)}\left(f_{N},f_{l},f_{t},f_{q}\right)=-\,\frac{\,e^{\myE_{l}+\myE_{t}}\;
      \left(-1+f_{N}+e^{\myE_{N}}\,f_{N}\right)}
    {\left(1+e^{\myE_{l}}\right)\,\left(1+e^{\myE_{t}}\right)\,
      \left(e^{\myE_{l}+\myE_{N}}+e^{\myE_{t}} \right)},
\end{equation}
where we have used energy conservation, and Fermi--Dirac statistics
for the leptons and quarks.

The integration over $q$ can now be performed analytically, reducing
the dimensions of the collision integrals to two.  These final
integrals must be evaluated numerically and then summed to give
$C_{S,s}[f_N]$,
\begin{equation}
 C_{S,s}[f_N] =
 C_{s}^{(1)}+C_{s}^{(2)}+C_{s}^{(3)}+C_{s}^{(4)}+C_{s}^{(5)}+C_{s}^{(6)}.
\end{equation}
The integrals $C_{S,s}^{(1,\ldots,6)}$ are as follows:

\begin{itemize}
\item The first integral comes from evaluating the $\Theta$ function
  $\Omega_{1a}$, and we have defined $\tilde{q}\equiv q/T$:

 \begin{align}
    \label{eq:Cs1}
    C_{S,s}^{(1)}&=\frac{3\,T}{2^{6}\pi^{3}\,\myE_{N}\,
      y_{N}}\,\frac{h_t^{2} \,M\,\tilde{m}_{1}}{v^{2}}
    \int_{y_{N}}^{\infty}\,
    d\myE_{l}\,\int_{0}^{\myE_{l}+\myE_{N}}\,d\myE_{t}\;\Lambda_{s}^{(N)}\;I_{s}^{(1)},
    \\ \nonumber
    I_{s}^{(1)}&=\int_{\myE_{l}-y_{N}}^{\myE_{l}+y_{N}}\,d \tilde{q}\,
    \frac{\left(\myE_{N}+\myE_{l}\right)^{2}-z^{2}-\tilde{q}^{2}
    }{\left(\myE_{N}+\myE_{l}\right)^{2}
      -\tilde{q}^{2}}  \\
    \label{eq:Is1} &= \frac{4\,y_{N}\,\left(\myE_{l}+\myE_{N}\right)
      + z^{2}\, \log
      \left[\frac{\left(\myE_{N}-y_{N}\right)\,\left(2\myE_{l}+\myE_{N}-y_{N}\right)}
        {\left(\myE_{N}+y_{N}\right)\,\left(2\myE_{l}+\myE_{N}+y_{N}\right)}
      \right] }{2\left(\myE_{l}+\myE_{N}\right)}.
  \end{align}

\item The second integral comes from evaluation of $\Omega_{1b}$:
  
  \begin{align}
    \label{eq:Cs1-b}
    C_{S,s}^{(2)}&=\frac{3\,T}{2^{6}\pi^{3}\,\myE_{N}\,
      y_{N}}\,\frac{h_t^{2} \,M\,\tilde{m}_{1}}{v^{2}}
    \int_{0}^{y_{N}}\,
    d\myE_{l}\,\int_{0}^{\myE_{l}+\myE_{N}}\,d\myE_{t}\;\Lambda_{s}^{(N)}\;I_{s}^{(2)},
    \\ \nonumber
    I_{s}^{(2)}&=\int_{y_{N}-\myE_{l}}^{\myE_{l}+y_{N}}\,d \tilde{q}\,
    \frac{\left(\myE_{N}+\myE_{l}\right)^{2}-z^{2}-\tilde{q}^{2}
    }{\left(\myE_{N}+\myE_{l}\right)^{2}
      -\tilde{q}^{2}}  \\
    \label{eq:Is1-b}
    &=\frac{4\,\myE_{l}\,\left(\myE_{l}+\myE_{N}\right) + z^{2}\, \log
      \left[\frac{\myE_{N}^{2}-y_{N}^{2}}
        {\left(2\myE_{l}+\myE_{N}\right)^{2}-y_{N}^{2}} \right]
    }{2\left(\myE_{l}+\myE_{N}\right)}.
  \end{align}

\item The third integral comes from the $\Omega_{2a,i}$ term: 

  \begin{align}
    \label{eq:Cs2}
    C_{S,s}^{(3)}&=\frac{3\,T}{2^{6}\pi^{3}\,\myE_{N}\, y_{N}
    }\,\frac{h_t^{2} \,M\,\tilde{m}_{1}}{v^{2}}
    \int_{y_N}^{\infty}\,
    d\myE_{l}\,\int_{0}^{\frac{1}{2}\,\left(\myE_{N}+y_{N}\right)}\,
    d\myE_{t}\;\Lambda_{s}^{(N)}\;I_{s}^{(3)}, \\ \nonumber
    I_{s}^{(3)}&=-\,\int_{\myE_{l}-y_{N}}^{\myE_{l}+\myE_{N}-2\myE_{t}}\,d \tilde{q} \,
 \frac{\left(\myE_{N}+\myE_{l}\right)^{2}-z^{2}-\tilde{q}^{2}
  }{\left(\myE_{N}+\myE_{l}\right)^{2}
      -\tilde{q}^{2}}    
\\ 
\label{eq:Is2} &=-\, 
    \frac{2\;\left(\myE_{l}+\myE_{N}\right)\, \left(\myE_{N}-2\,
        \myE_{t}+y_{N}\right) + z^{2}\, \log \left[\frac{\myE_{t}\,
          \left( 2\,\myE_{l}+\myE_{N}-y_{N}\right)}
        {\left(\myE_{l}+\myE_{N}-\myE_{t}\right)\,
          \left(\myE_{N}+y_{N}\right)}
      \right]}{2 \left(\myE_{l}+\myE_{N}\right)}.
  \end{align}
  
\item Integral four originates from the $\Omega_{2a,ii}$ term: 
  \begin{align}
    \label{eq:Cs3}
    C_{S,s}^{(4)}&=\frac{3\,T}{2^{6}\pi^{3}\,\myE_{N}\, y_{N}
    }\,\frac{h_t^{2} \,M\,\tilde{m}_{1}}{v^{2}}
    \int_{y_N}^{\infty}\,
    d\myE_{l}\,\int_{\frac{1}{2}\,\left(2\,\myE_{l}+\myE_{N}-y_{N}
      \right)}^{\myE_{l}+\myE_{N}}\,d\myE_{t}\;\Lambda_{s}^{(N)}\;I_{s}^{(4)}, \\
    \nonumber
    I_{s}^{(4)}&=-\,\int_{\myE_{l}-y_{N}}^{2\myE_{t}-\myE_{l}-\myE_{N}}\,
d \tilde{q} \,
 \frac{\left(\myE_{N}+\myE_{l}\right)^{2}-z^{2}-\tilde{q}^{2}
  }{\left(\myE_{N}+\myE_{l}\right)^{2}
      -\tilde{q}^{2}}    
\\ 
\label{eq:Is3} &= \frac{
      2\;\left(\myE_{l}+\myE_{N}\right)\,
      \left(2\,\myE_{l}+\myE_{N}-2\, \myE_{t}-y_{N}\right) - z^{2}\,
      \log \left[\frac{ \left(\myE_{l}+\myE_{N}-\myE_{t}\right)\,
          \left( 2\,\myE_{l}+\myE_{N}-y_{N}\right)}{\myE_{t}\,
          \left(\myE_{N}+y_{N}\right) }
      \right]}{2 \left(\myE_{l}+\myE_{N}\right)}.
 \end{align}
  
\item Integral five relates to $\Omega_{2b,i}$: 
\begin{align}
  \label{eq:Cs4}
  C_{S,s}^{(5)}&=\frac{3\,T}{2^{6}\pi^{3}\,\myE_{N}\, y_{N}
  }\,\frac{h_t^{2} \,M\,\tilde{m}_{1}}{v^{2}} \int_{0}^{y_N}\,
  d\myE_{l}\,\int_{0}^{\frac{1}{2}\,\left(2\,\myE_{l}+\myE_{N}-y_{N}
    \right)}\,d\myE_{t}\;\Lambda_{s}^{(N)}\;I_{s}^{(5)},\\ \nonumber
  I_{s}^{(5)}&=-\,\int_{y_{N}-\myE_{l}}^{\myE_{l}+\myE_{N}-2\myE_{t}}\,
d \tilde{q} \,
 \frac{\left(\myE_{N}+\myE_{l}\right)^{2}-z^{2}-\tilde{q}^{2}
  }{\left(\myE_{N}+\myE_{l}\right)^{2}
      -\tilde{q}^{2}}   
\\ 
\label{eq:Is4} &=-\,
  \frac{2\;\left(\myE_{l}+\myE_{N}\right)\,
    \left(2\,\myE_{l}+\myE_{N}-2\, \myE_{t}-y_{N}\right) - z^{2}\,
    \log \left[\frac{ \left(\myE_{l}+\myE_{N}-\myE_{t}\right)\, \left(
          2\,\myE_{l}+\myE_{N}-y_{N}\right)}{\myE_{t}\,
        \left(\myE_{N}+y_{N}\right) }
    \right]}{2\left(\myE_{l}+\myE_{N}\right)}.
  \end{align}

\item Finally, the sixth integral derives from $\Omega_{2b,ii}$:
  \begin{align}
    \label{eq:Cs5}
    C_{S,s}^{(6)}&=\frac{3\,T}{2^{6}\pi^{3}\,\myE_{N}\, y_{N}
    }\,\frac{h_t^{2} \,M\,\tilde{m}_{1}}{v^{2}}
    \int_{0}^{y_N}\,
    d\myE_{l}\,\int_{\frac{1}{2}\,\left(\myE_{N}+y_{N}\right)}^{\myE_{l}+\myE_{N}}\,
    d\myE_{t}\;\Lambda_{s}^{(N)}\;I_{s}^{(6)},\\
    \nonumber
    I_{s}^{(6)}&=-\,\int_{y_{N}-\myE_{l}}^{2\myE_{t}-\myE_{l}-\myE_{N}}\,
d \tilde{q} \,
 \frac{\left(\myE_{N}+\myE_{l}\right)^{2}-z^{2}-\tilde{q}^{2}
  }{\left(\myE_{N}+\myE_{l}\right)^{2}
      -\tilde{q}^{2}}   
\\ 
\label{eq:Is5} &=
    \frac{2\;\left(\myE_{l}+\myE_{N}\right)\, \left(\myE_{N}-2\,
        \myE_{t}+y_{N}\right) - z^{2}\, \log \left[\frac{
          \left(\myE_{l}+\myE_{N}-\myE_{t}\right)\, \left(
            \myE_{N}+y_{N}\right)}{\myE_{t}\,\left(2\,\myE_{l}+
            \myE_{N}-y_{N}\right) }
      \right]}{2 \left(\myE_{l}+\myE_{N}\right)}.
  \end{align}

\end{itemize}

\subsubsection{Lepton asymmetry}
\label{sec:lepton-asymmetry-s}

The $s$-channel collision term in Eq.~(\ref{eq:be-asy-fund-scat}) for
tracking the lepton asymmetry is
\begin{align}
  \label{eq:C-s-l-1}
  C_{S,s}[f_{l-\overline{l}}] =\frac{1}{2E_{l}}\int \prod_{i=N,q,t}
  \frac{dp^{3}_{i}}{(2\pi)^{3}\,2E_{i}}(2\pi)^{4}
  \delta^{4}(p_{l}+p_{N}-p_{q}-p_{t}) \vert
  \mathcal{M}_{s}\vert^{2}\Lambda_{s}^{(l-\overline{l})}
  \left(f_{l-\overline{l}},f_{N},f_{t},t_{q}\right),
\end{align}
with phase space factor
\begin{equation}
  \label{eq:lam-s-a-1}
  \Lambda_{s}^{(l-\overline{l})}
  \left(f_{l-\overline{l}},f_{N},f_{t},t_{q}\right)=  
  f_{l-\overline{l}}\,\left(f_{N}\,\left(f_{t}+f_{q}-1\right)-
    f_{t}f_{q} \right),
\end{equation}
and matrix element $\mathcal{M}_s$ given by Eq.~(\ref{eq:s-n-matrix-3}).

Again we choose an explicit coordinate system and use rotational
invariance
\begin{equation}
\begin{aligned}
  \label{eq:C-s-l-5}
  \mathbf{q} &= \vert \mathbf{q} \vert \,(0,0,1), \\
  \mathbf{p}_{l} &= E_{l} \, (0, \sin \eta , \cos \eta  ), \\
  \mathbf{p}_{t} &= E_{t} (\cos \phi\, \sin \vartheta, \sin \phi \,
  \sin \vartheta , \cos \vartheta).
\end{aligned}
\end{equation}
Following the same procedure as in section~\ref{sec:right-hand-neutr},
we reduce the collision integral~(\ref{eq:C-s-l-1}) to
\begin{align}
  \label{eq:C-t-l-c1}
  C_{S,s}[f_{l-\overline{l}}]&= \frac{1}{2^{7}\pi^{3}\,E^{2}_{l}} \int
  d\cos\vartheta \, d\cos\eta \, dE_{t}\, dE_{N}\, dq \,\vert
  \mathcal{M}_{s} \vert^{2} \\ \nonumber & \times \delta \left( \!
    \cos\eta \!-\frac{E^{2}_{l}-E^{2}_{N}+ M +\vert \mathbf{q}
      \vert^{2}}{2 \vert \mathbf{q} \vert E_{l}}\right) \delta
  \left(\! \cos\vartheta\! - \frac{E^{2}_{t}-\left( E_{l}+E_{N}-E_{t}
      \right)^{2} +\vert \mathbf{q} \vert^{2} }{2\vert \mathbf{q}\vert
      E_{t}} \right ) \\ \nonumber & \times
  \Lambda_{s}^{(l-\overline{l})}
  \left(f_{l-\overline{l}},f_{N},f_{t},t_{q}\right) 
\Theta\left(E_{N}\right)\,\Theta \left(E_{t}\right) \,
\Theta
  \left(E_{l}+E_{N}-E_{t}\right),
\end{align}
with phase space element
\begin{equation}
  \label{eq:lam-s-a-2}
  \Lambda_{s}^{(l-\overline{l})}
  \left(f_{l-\overline{l}},f_{N},f_{t},t_{q}\right)= -\,f_{l-\overline{l}}\,
  \frac{e^{\myE_{t}}\,\left(1+\left(e^{\myE_{l}+\myE_{N}}-1\right)\,f_{N}\right)}
  {\left(1+e^{\myE_{t}}\right)\,\left(e^{\myE_{l}+\myE_{N}}+e^{\myE_{t}}\right)},  
\end{equation}
using as usual energy conservation, and Fermi--Dirac statistics for
the leptons and quarks.

In analogy to section~\ref{sec:right-hand-neutr}, we further reduce
the collision integral~(\ref{eq:C-t-l-c1}) to a sum of six integrals
with distinct integration ranges, 
\begin{equation}
  C_{S,s}[f_{l-\overline{l}}] =  C_{S,s}^{(1)}+C_{S,s}^{(2)}+C_{S,s}^{(3)}+C_{S,s}^{(4)}+C_{S,s}^{(5)}+C_{S,s}^{(6)},
\end{equation}
to be integrated numerically over two remaining degrees of freedom.
These integrals are:
\begin{itemize}
\item First integral 
\begin{align}
  \label{eq:Cas1}
  C_{S,s}^{(1)}&= \frac{3\,T}{2^{6}\pi^{3}\,
    \myE^{2}_{l}}\,\frac{h_t^{2} \,M\,\tilde{m}_{1}}{v^{2}}
  \int_{z}^{\sqrt{\myE^2_l+z^2}}\, d
  \myE_{N}\,\int_{0}^{\myE_{l}+\myE_{N}}\,d\myE_{t}\;
  \Lambda_{s}^{(l-\overline{l})}\; I_{s}^{(1)},
  \end{align}
where $I_{s}^{(1)}$ is given by Eq.~(\ref{eq:Is1}).

\item Second integral with $I_{s}^{(2)}$ given by 
  Eq.~(\ref{eq:Is1-b}):
   \begin{align}
    \label{eq:Cas1a}
    C_{S,s}^{(2)}&= \frac{3\,T}{2^{6}\pi^{3}\,
      \myE^{2}_{l}}\,\frac{h_t^{2} \,M\,\tilde{m}_{1}}{v^{2}}
    \int_{\sqrt{\myE_l^2+z^2}}^{\infty}\, d
    \myE_{N}\,\int_{0}^{\myE_{l}+\myE_{N}}\,d\myE_{t}\;
    \Lambda_{s}^{(l-\overline{l})}\; I_{s}^{(2)}.
  \end{align}

\item Third integral ($I_s^{(3)}$ given by Eq.~(\ref{eq:Is2})):  
  \begin{align}
    \label{eq:Cas2}
    C_{S,s}^{(3)}&= \frac{3\,T}{2^{6}\pi^{3}\,
      \myE^{2}_{l}}\,\frac{h_t^{2} \,M\,\tilde{m}_{1}}{v^{2}}
    \int_{z}^{\sqrt{\myE_l^2+z^2}}\, d
    \myE_{N}\,\int_{0}^{\frac{1}{2}\,\left(\myE_{N}+y_{N}\right)}\,
    d\myE_{t}\;\Lambda_{s}^{(l-\overline{l})}\; I_{s}^{(3)}.
  \end{align}

\item Fourth integral ($I_s^{(4)}$ given by Eq.~(\ref{eq:Is3})): 
  \begin{align}
    \label{eq:Cas3}
    C_{S,s}^{(4)}&= \frac{3\,T}{2^{6}\pi^{3}\,
      \myE^{2}_{l}}\,\frac{h_t^{2} \,M\,\tilde{m}_{1}}{v^{2}}
    \int_{z}^{\sqrt{\myE_l^2+z^2}}\, d \myE_{N}\,
    \int_{\frac{1}{2}\,\left(2\,\myE_{l}+\myE_{N}-y_{N}\right)}^{\myE_{l}+\myE_{N}}\,
    d\myE_{t}\;\Lambda_{s}^{(l-\overline{l})}\; I_{s}^{(4)}.
  \end{align}

\item Fifth integral ($I_s^{(5)}$ given by Eq.~(\ref{eq:Is4})): 
  \begin{align}
    \label{eq:Cas4}
    C_{S,s}^{(5)}&= \frac{3\,T}{2^{6}\pi^{3}\,
      \myE^{2}_{l}}\,\frac{h_t^{2} \,M\,\tilde{m}_{1}}{v^{2}}
    \int_{\sqrt{\myE_l^2 + z^2}}^{\infty}\, d
    \myE_{N}\,\int_{0}^{\frac{1}{2}\,\left(2\,\myE_{l}+\
        \myE_{N}-y_{N}\right)}\,d\myE_{t}\;\Lambda_{s}^{(l-\overline{l})}\;
    I_{s}^{(5)}.
  \end{align}

\item sixth integral ($I_s^{(6)}$ given by Eq.~(\ref{eq:Is5})): 
  \begin{align}
    \label{eq:Cas5}
    C_{S,s}^{(6)}&= \frac{3\,T}{2^{6}\pi^{3}\,
      \myE^{2}_{l}}\,\frac{h_t^{2} \,M\,\tilde{m}_{1}}{v^{2}}
    \int_{\sqrt{\myE^2_l + z^2}}^{\infty}\, d
    \myE_{N}\,\int_{\frac{1}{2}\,\left(\myE_{N}+y_{N}\right)}
    ^{\myE_{l}+\myE_{N}}\,d\myE_{t}\;\Lambda_{s}^{(l-\overline{l})}\;
    I_{s}^{(6)}.
  \end{align}

\end{itemize}

\subsection{$t$-channel}
\label{subsec:full-t-channel}

\subsubsection{Right-handed neutrino}\label{sec:right-hand-neutr-t}

The collision integral for the $t$-channel process appearing 
in Eq.~(\ref{eq:be-rhn-fund-scat}) is given by
\begin{align}
  \label{eq:C-t-n-1}
  C_{S,t}[f_N] &= \frac{1}{2E_{N}} \, \int \prod_{i=l,q,t}
  \frac{dp^{3}_{i}}{(2\pi)^{3}2E_{i}} (2\pi)^{4} \delta^{4}
  (p_{N}+p_{q}- p_{t}-p_{l}) \, \vert \mathcal{M}_{t} \vert^{2}\;
  \Lambda_t^{(N)}\left(f_{N},f_{q},f_{l},f_{t}\right),
\end{align}
with phase space factor
\begin{equation}
  \label{eq:lam-t-n-1}
  \Lambda_t^{(N)}\left(f_{N},f_{q},f_{l},f_{t}\right)=
  \left[ (1-f_{N})(1-f_{q})f_{t}f_{l} -
    f_{N}f_{q}(1-f_{t})(1-f_{l}) \right],
\end{equation}
and 
\begin{equation}
  \label{eq:C-t-n-1a}
  \vert \mathcal{M}_{t} \vert^{2}=24\,h_t^{2}
  \frac{M\tilde{m}_{1}}{v^{2}} \, \frac{p_{N}p_{l}\,p_{q}p_{t}}{t^{2}}
\end{equation}
is the matrix element.

The reduction of the collision integral proceeds in the same way as
for the analogous $s$-channel collision integral in
section~\ref{subsec:full-s-channel}.  We use the momentum
\begin{equation}
  \label{eq:C-t-n-2}
  \mathbf{k} \equiv \mathbf{p}_{l}-\mathbf{p}_{N}=
  \mathbf{p}_{q}-\mathbf{p}_{t},
\end{equation}
and coordinates
\begin{equation}
  \label{eq:C-t-n-5}
  \begin{aligned}
    \mathbf{k} &= \vert \mathbf{k}\vert\,(0,0,1), \\
    \mathbf{p}_{q} &= E_{q} \, (0, \sin \eta , \cos \eta  ), \\
    \mathbf{p}_{N} &= \vert \mathbf{p}_{N}\vert (\cos
    \phi\, \sin \vartheta, \sin \phi \, \sin \vartheta ,\cos \vartheta).
  \end{aligned}
\end{equation}
The matrix element in these coordinates reads 
\begin{equation}
  \label{eq:t-n-matrix-3}
  \vert \mathcal{M}_{t} \vert ^{2}=  6\,h_t^{2}
  \frac{M\tilde{m}_{1}}{v^{2}} \,
  \frac{\left(E_{N}-E_{l}\right)^{2}-M^{2}-
    \vert\mathbf{k}\vert^{2}
  }{\left(E_{N}-E_{l}\right)^{2}
    -\vert\mathbf{k}\vert^{2}}.
\end{equation}
Averaging over the incoming RHN direction and integrating over all
azimuthal angles leads to the following 5-dimensional integral: 
\begin{equation}
\begin{aligned}
  \label{eq:C-t-n-8}
  C_{S,t}[f_N]&= \frac{1}{2^{7}\pi^{3}\,E_{N}\,p_{N}} \int
  d\cos \vartheta \, d\cos \eta \, dE_{q}\, dE_{l}\,
  d\vert \mathbf{k}\vert \, \vert \mathcal{M}_{t} \vert^{2} \\
  &\times \delta \left(
    \cos\eta \, + \frac{\left(E_{N}+E_{q}-E_{l}\right)^{2} -E^{2}_{q}
      -\vert \mathbf{k}^{2}\vert}{2\vert \mathbf{k}\vert\,E_{q}}
  \right) \delta \left( \cos\vartheta-
    \frac{E^{2}_{l}-E^{2}_{N}+M-\vert \mathbf{k}^{2}\vert}{2
      \vert \mathbf{k}\vert\, \vert \mathbf{p}_{N}\vert} \right)  \\
  &\times \Lambda_t^{(N)}\left(f_{N},f_{q},f_{l},f_{t}\right)\,
  \Theta\left(E_{q}\right)\, \Theta \left(E_{l}\right)\,
\Theta
  \left(E_{N}+E_{q}-E_{l}\right),
\end{aligned}
\end{equation}
with
\begin{equation}
  \label{eq:lam-t-n-2}
  \Lambda_t^{(N)}\left(f_{N},f_{q},f_{l},f_{t}\right)=
  - \frac{
    e^{\myE_{l}+\myE_{q}}\, \left(-1+f_{N}+e^{\myE_{N}}\,f_{N}
    \right)}{\left(1+e^{\myE_{l}}\right)\,
    \left(1+e^{\myE_{q}}\right)\,\left(e^{\myE_{l}}+e^{\myE_{l}+\myE_{q}}\right)},
\end{equation}
assuming thermal equilibrium for the standard model particles.

The integrals over $\cos \eta$ and $\cos \vartheta$ in
Eq.~(\ref{eq:C-t-n-8}) can be readily performed in the same manner as
before, and, in the process, integration limits are derived for the
integral over $k=\vert \mathbf{k} \vert$.  As it turns out, there are
two possible lower limits for the $k$-integration: $k_{{\rm
    min},1}=E_{N}-E_{l}$ and $k_{{\rm min},2}=E_{l}-p_{N}$.  Here, a
comment on the infrared cut-off is in order.  In the integrated
picture, a divergence occurs in the integral over $t$ at $t=0$, whose
regulation requires the introduction of a Higgs mass in the
propagator, i.e., $\vert \mathcal{M}_{t} \vert^2 \propto 1/t \to \vert
\mathcal{M}_{t} \vert^2 \propto 1/(t-m^{2}_{\Phi})$.  In the full
treatment, the matrix element has the form $\vert \mathcal{M}_{t}
\vert^2 \propto 1/(\left ( E_{N}-E_{l} \right)^{2} -\vert k^{2})$, so
that the equivalent divergence occurs in the integration over $k$ at
$k=E_{N}-E_{l}$, i.e., at $k=k_{{\rm min},1}$.  This divergence can be
avoided simply by modifying by hand the integration limit
$\int_{k_{{\rm min},1}}\rightarrow \int_{k_{{\rm min},1}+m_{\Phi}}$.
There are no changes for those integrals with $k_{\rm min}=k_{{\rm
    min},2}$.

It is also possible to regulate the divergence by introducing a Higgs
mass in the propagator, such as in the integrated picture.  This
modifies the integration over $k$ not only for $E_{N}>E_{l}$ (i.e.,
$k_{\rm min}=k_{{\rm min},1}$), but also for $E_{l}>p_{N}$ (i.e.,
$k_{\rm min}=k_{{\rm min},2}$).  In physical terms this procedure
corresponds to giving the Higgs particle a mass whose magnitude can
vary from zero up to possible thermal contributions, i.e., $0 \leq
m_{\Phi}/M \lesssim 0.4\,T/M$, where $m_{\Phi}(T)\sim 0.4\,T$ is the
thermal Higgs mass~\cite{Giudice:2003jh}.  In the temperature regime
relevant to leptogenesis, electroweak symmetry is unbroken and
therefore leptons are massless.

Since we have so far not included thermal corrections in the present
work, for consistency we prefer not to use the thermal Higgs mass.
Furthermore, in a full thermal treatment, RHN decay into a lepton and
Higgs pairs become kinematically forbidden at high enough
temperatures, and the decay of a Higgs particle into a neutrino and
lepton pair becomes viable~\cite{Giudice:2003jh}. Thus, in addition to
determining the value of the infrared cut-off there is also the
question of its interpretation.  In view of these issues, we choose to
deal with the infrared divergence using the simpler method of cutting
off the integration over $k$ at $k_{\rm min}=k_{{\rm min},1}+m_\Phi$,
with $a_h=m_{\Phi}/M=10^{-5}$.

After integrating over $\cos \eta$ and $\cos \vartheta$, 
the original integral~(\ref{eq:C-t-n-8}) is now split into four parts
\begin{equation}
  C_{S,t}[f_N]= C_{S,t}^{(1)} + C_{S,t}^{(2)} +C_{S,t}^{(3)} +C_{S,t}^{(4)},
\end{equation}
where the constituent integrals are as follows:
\begin{itemize}
\item First integral (with $\tilde{k}\equiv k/T$):
\begin{align}
  \label{eq:C-t-n-10}
  C_{S,t}^{(1)}&=\frac{3\,T}{2^{6}\pi^{3}\, \myE_{N}
    y_{N}}\,\frac{h_t^{2} \,M\,\tilde{m}_{1}}{v^{2}}
  \int_{\frac{1}{2}\left(\myE_{N}-y_{N}+a_h z\right)}^{\frac{1}{2}\left(\myE_{N}+y_{N}\right)}
  d\myE_{l} \int_{\frac{1}{2} a_h z}
  ^{\frac{1}{2}\left(2\myE_{l}-\myE_{N}+y_{N}\right)} d\myE_{q}\;
  \Lambda_t^{(N)} \; I_{t}^{(1)}, \\ 
\nonumber I_{t}^{(1)} &=
  \int_{\myE_{N}-\myE_{l}+a_h z}^{2\myE_{q}+\myE_{N}-\myE_{l}}\,
d \tilde{k} \,
 \frac{\left(\myE_{N}-\myE_{l}\right)^{2}-z^{2}-\tilde{k}^{2}
  }{\left(\myE_{N}-\myE_{l}\right)^{2}
      -\tilde{k}^{2}}   
\\ \nonumber & = \frac{ 2
    \left(\myE_{N}-\myE_{l}\right)\, \left(2\myE_{q}-a_{h}z \right)-
    z^{2}\,
    \log\left[\frac{\left(\myE_{N}+\myE_{q}-\myE_{l}\right)\,a_{h}z}
      {\myE_{q}\left(2\left(\myE_{N}-\myE_{l}\right)+a_{h}z\right)}
    \right]} {2\, \left(\myE_{N}-\myE_{l}\right)}.
   \end{align}

\item Second integral:
\begin{align}
  \label{eq:C-t-n-12}
  C_{S,t}^{(2)}&=\frac{3\,T}{2^{6}\pi^{3}\, \myE_{N}
    y_{N}}\,\frac{h_t^{2} \,M\,\tilde{m}_{1}}{v^{2}}
  \int_{\frac{1}{2}\left(\myE_{N}-y_{N} + a_h z\right)}^{\frac{1}{2}\left(\myE_{N}+y_{N}\right)}
  d\myE_{l} \int_{\frac{1}{2}(2\myE_{l}-\myE_{N}+y_{N})}^{\infty}
  d\myE_{q} \;\Lambda_t^{(N)} \;I_{t}^{(2)}, \\ \nonumber
  I_{t}^{(2)}&= \int_{\myE_{N}-\myE_{l}+a_h z}^{\myE_{l}+y_{N}}\,
d \tilde{k} \,
 \frac{\left(\myE_{N}-\myE_{l}\right)^{2}-z^{2}-\tilde{k}^{2}
  }{\left(\myE_{N}-\myE_{l}\right)^{2}
      -\tilde{k}^{2}}   
\\ \nonumber & =  \frac{2
    \,\left(\myE_{N}-\myE_{l}\right)\,\left(2
      \myE_{l}-\myE_{N}+y_{N}-a_{h}z\right) -z^{2}
    \,\left(\log\left[\frac{-a_{h}z\,\left(\myE_{N}+y_{N}\right)}
        {\left(\myE_{N}-2\myE_{l}-y_{N}\right)\,
          \left(2\left(\myE_{N}-\myE_{l}\right)+a_{h}z\right)} \right]
    \right)} {2\,\left(\myE_{N}-\myE_{l}\right)}.
  \end{align}

\item Third integral:
  \begin{align}
    \label{eq:C-t-n-14}
    C_{S,t}^{(3)}&=\frac{3\,T}{2^{6}\pi^{3}\, \myE_{N}
      y_{N}}\,\frac{h_t^{2} \,M\,\tilde{m}_{1}}{v^{2}}
    \int_{\frac{1}{2} \left(\myE_N+y_N\right)}^{\infty} d\myE_{l}
    \int_{\frac{1}{2}\left(2\myE_{l}-\myE_{N}-y_{N} \right)}
    ^{\frac{1}{2}\left(2\myE_{l}-\myE_{N}+y_{N}\right)}
    d\myE_{q}\;\Lambda_t^{(N)} \;I_{t}^{(3)}, \\ \nonumber
    I_{t}^{(3)}&=
    \int_{\myE_{l}-y_{N}}^{2\myE_{q}+\myE_{N}-\myE_{l}}\,
d \tilde{k} \,
 \frac{\left(\myE_{N}-\myE_{l}\right)^{2}-z^{2}-\tilde{k}^{2}
  }{\left(\myE_{N}-\myE_{l}\right)^{2}
      -\tilde{k}^{2}}   
\\ \nonumber & = \frac{2
      \left(\myE_{N}-\myE_{l}\right)\,\left(\myE_{N}+y_{N}+
        2\left(\myE_{q}-\myE_{l}\right) \right) +z^{2}\,\left(
        \log\left[ -\frac{-\myE_{q}\left(\myE_{N}-y_{N}\right)}
          {\left(\myE_{N}+\myE_{q}-\myE_{l}\right)\,
            \left(\myE_{N}-2\myE_{l}+y_{N}\right)}\right]\right)
    }{2\,\left(\myE_{N}-\myE_{l}\right)}.
  \end{align}

\item Fourth integral:
  \begin{align}
    \label{eq:C-t-n-16}
    C_{S,t}^{(4)}&= \frac{3\,T}{2^{6}\pi^{3}\, \myE_{N}
      y_{N}}\,\frac{h_t^{2} \,M\,\tilde{m}_{1}}{v^{2}}
    \int_{\frac{1}{2}\left(\myE_{N}+y_{N}\right)}^{\infty} d\myE_{l}
    \int_{\frac{1}{2}\left(\myE_{l}-\myE_{N}+y_{N}\right)}^{\infty}
    d\myE_{q}\;\Lambda_t^{(N)} \;I_{t}^{(4)},\\ \nonumber
    I_{t}^{(4)} &=
    \int_{\myE_{l}-y_{N}}^{\myE_{l}+y_{N}}\,
d \tilde{k} \,
 \frac{\left(\myE_{N}-\myE_{l}\right)^{2}-z^{2}-\tilde{k}^{2}
  }{\left(\myE_{N}-\myE_{l}\right)^{2}
      -\tilde{k}^{2}}   
\\
    \nonumber & = 
    \frac{4\,\left(\myE_{N}-\myE_{l}\right)\,y_{N}+z^{2}\, \log\left[
        \frac{\left(\myE_{N}-y_{N}\right)\,
          \left(\myE_{N}-y_{N}-2\myE_{l}\right)}{\left(\myE_{N}+y_{N}\right)\,
          \left(\myE_{N}+y_{N}-2\myE_{l}\right)} \right]}
    {2\,\left(\myE_{N}-\myE_{l}\right)}.
  \end{align}

\end{itemize}

\subsubsection{Lepton asymmetry}\label{sec:lepton-asymmetry-t}

The $t$-channel collision integral for the lepton asymmetry evolution
is
\begin{align}
  \label{eq:C-t-l-0}
  C_{S,t}[f_{l-\overline{l}}]& =\frac{1}{2E_{l}}\int \prod_{i=N,q,t}
  \frac{dp^{3}_{i}}{(2\pi)^{3}\,2E_{i}}(2\pi)^{4}
  \delta^{4}(p_{l}+p_{q}-p_{N}-p_{t}) \vert
  \mathcal{M}_{t}\vert^{2}\Lambda_t^{(l-\overline{l})}
  \left(f_{l-\overline{l}},f_{t},f_{N},f_{q}\right),
  \end{align}
with
\begin{equation}
  \label{eq:lam-t-a-1}
\Lambda_t^{(l-\overline{l})}
  \left(f_{l-\overline{l}},f_{t},f_{N},f_{q}\right)= 
  f_{l-\overline{l}}\,\left(f_{q}\,\left(f_{t}+f_{N}-1\right)-
    f_{t}f_{N} \right).
\end{equation}
The matrix element is the same as for the RHN given in
Eq.~(\ref{eq:C-t-n-1a}), and 
we use the momentum
\begin{equation}
  \label{eq:C-t-l-1}
  \mathbf{k}\equiv \mathbf{p}_{N}-\mathbf{p}_{l}=
  \mathbf{p}_{q}-\mathbf{p}_{t},
\end{equation}
and coordinates
\begin{align}
  \label{eq:C-t-l-4}
    \mathbf{k} &= \vert \mathbf{k}\vert\,(0,0,1), \\ \nonumber
    \mathbf{p}_{q} &= E_{q} \, (0, \sin \eta , \cos \eta ) , \\ \nonumber
    \mathbf{p}_{l} &= E_{l}\, (\cos \phi\, \sin \vartheta, \sin
    \phi \, \sin \vartheta , \cos \vartheta).
  \end{align}
  Following the method of the previous sections, we reduce the
  integral~(\ref{eq:C-t-l-0}) to 
\begin{equation}
\begin{aligned}
  \label{eq:C-t-l-7}
  C_{S,t}[f_{l-\overline{l}}]&= \frac{1}{2^{7}\pi^{3}\,E^{2}_{l}} \int d\cos \vartheta\,
  d\cos \eta \, dE_{q}\, dE_{N}\, d \vert \mathbf{k} \vert \, \vert \mathcal{M}_{t}
  \vert^{2} \\  &\times \delta \left(
    \cos\eta+\frac{\left(E_{l}+E_{q}-E_{N}\right)^{2} -E^{2}_{q}
      -\vert \mathbf{k}^{2}\vert}{2\vert \mathbf{k}\vert\,E_{q}}
  \right) \delta \left( \cos\vartheta-
    \frac{E^{2}_{N}-E^{2}_{l}-M^2-\vert \mathbf{k}^{2}\vert}{2
      \vert \mathbf{k}\vert\,E_{l}} \right)  \\
  & \times \Lambda_t^{(l-\overline{l})}
  \left(f_{l-\overline{l}},f_{t},f_{N},f_{q}\right)\,
  \Theta\left(E_{q}\right)\,\Theta\left(E_{N}\right)\,
  \Theta\left(E_{l}+E_{q}-E_{N}\right),
  \end{aligned}
\end{equation}
with
\begin{equation}
  \label{eq:lam-t-a-2}
  \Lambda_t^{(l-\overline{l})}
  \left(f_{l-\overline{l}},f_{t},f_{N},f_{q}\right)= f_{l-\overline{l}}\,
  \frac{e^{\myE_{q}}\left(e^{\myE_{l}}\,\left(-1+f_{N}\right)
      -e^{\myE_{N}}\,f_{N}\right)}
  {\left(1+e^{\myE_{q}}\right)
    \,\left(e^{\myE_{N}}+e^{\myE_{q}+\myE_{l}}\right)}
\end{equation}
as the phase space factor.

Integrating over $\cos \eta$ and $\cos \vartheta$, we 
split up the integral~(\ref{eq:C-t-l-7}) into four parts,
\begin{equation}
 C_{S,t}[f_{l-\overline{l}}]= C_{S,t}^{(1)} + C_{S,t}^{(2)} +C_{S,t}^{(3)} +C_{S,t}^{(4)},
\end{equation}
with:
\begin{itemize}
\item First integral: 
  \begin{align}
    \label{eq:C-t-l-9}
    C_{S,t}^{(1)}&=\frac{3\,T}{2^{6}\pi^{3}\, \myE_{N}
      y_{N}}\,\frac{h_t^{2} \,M\,\tilde{m}_{1}}{v^{2}}
    \int_{\frac{(2\myE_{l}-a_h z)^{2}+z^{2}}{2 (2 \myE_{l}-a_h
        z)}}^{\infty} d\myE_{N} \int_{\myE_{N}-\myE_{l}+\frac{1}{2}
      a_h z} ^{\frac{1}{2}\left(\myE_{N}+y_{N}\right)}
    d\myE_{q}\; \Lambda_t^{(l-\overline{l})}\;I_{t}^{(1)}, \\
    \nonumber I_{t}^{(1)} &=\int_{\myE_{N}-\myE_{l}+a_h
      z}^{2\myE_{q}+\myE_{l}-\myE_{N}}\, d \tilde{k} \,
    \frac{\left(\myE_{N}-\myE_{l}\right)^{2}-z^{2}-\tilde{k}^{2}
    }{\left(\myE_{N}-\myE_{l}\right)^{2} -\tilde{k}^{2}}
    \\
    \nonumber & = -\, \frac{ 2\,\left(\myE_{N}-\myE_{l}\right)\,
      \left(2\,\left(\myE_{N}-\myE_{q}-\myE_{l}\right)+a_{h}z\right) +
      z^{2} \,\log
      \left[\frac{\myE_{q}\,a_{h}z}{\left(\myE_{q}-\myE_{N}
            +\myE_{l}\right)\,\left(
            2\left(\myE_{N}-\myE_{l}\right)+a_{h}z\right)} \right]
    }{2\,\left(\myE_{N}-\myE_{l}\right)}.
  \end{align}

\item Second integral:
  \begin{align}
    \label{eq:C-t-l-11}
    C_{S,t}^{(2)}&=\frac{3\,T}{2^{6}\pi^{3}\, \myE_{N}
      y_{N}}\,\frac{h_t^{2} \,M\,\tilde{m}_{1}}{v^{2}}
    \int_{\frac{(2\myE_{l}-a_h z)^{2}+z^{2}}{2 (2 \myE_{l}-a_h
        z)}}^{\infty} d\myE_{N}
    \int_{\frac{1}{2}\left(\myE_{N}+y_{N}\right)}^{\infty} d\myE_{q}
    \;\Lambda_t^{(l-\overline{l})}\;I_{t}^{(2)}, \\ \nonumber
    I_{t}^{(2)}&=\int_{\myE_{N}-\myE_{l}+a_h z}^{\myE_{l}+y_{N}}\, d
    \tilde{k} \,
    \frac{\left(\myE_{N}-\myE_{l}\right)^{2}-z^{2}-\tilde{k}^{2}
    }{\left(\myE_{N}-\myE_{l}\right)^{2} -\tilde{k}^{2}}
    \\
    \nonumber & = \frac{ 2\,\left(\myE_{N}-\myE_{l}\right)\,
      \left(2\myE_{l}-\myE_{N}+y_{N}-a_{h}z\right) - z^{2} \,\log
      \left[\frac{\left(\myE_{N}+y_{N}\right)\,a_{h}z}
        {\left(2\myE_{l}-\myE_{N}+y_{N}\right)\,\left(
            2\left(\myE_{N}-\myE_{l}\right)+a_{h}z\right)} \right]
    }{2\,\left(\myE_{N}-\myE_{l}\right)}.
  \end{align}

\item Third integral: 
  \begin{align}
    \label{eq:C-t-l-13}
    C_{S,t}^{(3)}&=\frac{3\,T}{2^{6}\pi^{3}\, \myE_{N}
      y_{N}}\,\frac{h_t^{2} \,M\,\tilde{m}_{1}}{v^{2}}
    \int_{z}^{\frac{4\myE_{l}^{2}+z^{2}}{4\myE_{l}}} d\myE_{N}
    \int_{\frac{1}{2}(\myE_{N}-y_{N})}^{\frac{1}{2}(\myE_{N}+y_{N})}
    d\myE_{q} \;\Lambda_t^{(l-\overline{l})}\;I_{t}^{(3)},
    \\\nonumber I_{t}^{(3)}&=\int_{\myE_{l}-y_{N}}^{2\myE_{q}+\myE_{l}-\myE_{N}}\, 
d \tilde{k} \,
 \frac{\left(\myE_{N}-\myE_{l}\right)^{2}-z^{2}-\tilde{k}^{2}
  }{\left(\myE_{N}-\myE_{l}\right)^{2}  -\tilde{k}^{2}}
\\
    \nonumber &  = \, \frac{
        2\,\left(\myE_{N}-\myE_{l}\right)\,\left(2\myE_{q}-\myE_{N}+y_{N}\right)
      \, + z^{2} \,\log
        \left[\frac{\left(\myE_{N}-\myE_{q}-\myE_{l}\right)\,
            \left(\myE_{N}-y_{N}
            \right)}{\myE_{q}\,\left(\myE_{N}-2\myE_{l}+y_{N} \right)}
        \right]}{2\,\left(\myE_{N}-\myE_{l}\right)}.
  \end{align}

\item Fourth integral: 
  \begin{align}
    \label{eq:C-t-l-15}
    C_{S,t}^{(4)}&=\frac{3\,T}{2^{6}\pi^{3}\, \myE_{N}
      y_{N}}\,\frac{h_t^{2} \,M\,\tilde{m}_{1}}{v^{2}}
    \int_{z}^{\frac{4\myE_{l}^{2}+z^{2}}{4\myE_{l}}} d\myE_{N}
    \int_{\frac{1}{2}(\myE_{N}+y_{N})}^{\infty} d\myE_{q}
    \;\Lambda_t^{(l-\overline{l})}\;I_{t}^{(4)}, \\ \nonumber
    I_{t}^{(4)} &=\int_{\myE_{l}-y_{N}}^{\myE_l + y_N}\, 
d \tilde{k} \,
 \frac{\left(\myE_{N}-\myE_{l}\right)^{2}-z^{2}-\tilde{k}^{2}
  }{\left(\myE_{N}-\myE_{l}\right)^{2}  -\tilde{k}^{2}}
\\
    \nonumber&  = \, \frac{
        \,4\,\left(\myE_{N}-\myE_{l}\right)\,y_{N} + z^{2} \,\log
        \left[\frac{\left(\myE_{N}-2\myE_{l}-y_{N}\right)\,\left(\myE_{N}-y_{N}
            \right)}{\left(\myE_{N}-2\myE_{l}+y_{N}
            \right)\,\left(\myE_{N}+y_{N}\right)} \right]
      }{2\,\left(\myE_{N}-\myE_{l}\right)}.
  \end{align}

\end{itemize}

\section{Evolution of the top Yukawa coupling}
\label{sec:RGE}
To determine the gauge and Yukawa couplings at some 
energy scale $\mu \equiv \log\left(T/m_{Z}\right)$, we use the
renormalisation group equation,
\begin{align}
  \label{eq:RGE-1}
  \frac{dg_{i}^{2}}{d\mu}=\frac{c_{1}}{8\pi^{2}}\,g_{i}^{4}.
\end{align}
where $i$ denotes the corresponding gauge group of the standard
model.  The constants for the gauge couplings are
\begin{align}
  \label{eq:beta-1}
  \left(c_{1},c_{2},c_{3}\right)=
  \left(\frac{41}{10}\cdot\frac{5}{3},\frac{16}{9},-7 \right).
\end{align}
At one loop the solution of~(\ref{eq:RGE-1}) for the gauge
couplings yields 
\begin{align}
  \label{eq:RGE-sol}
  g_{i}=\sqrt{\frac{g_{i}^2(\mu=0)}{1-\frac{g_{i}^2(\mu=0)}{8\pi^{2}}\,\mu}}.
\end{align}
Neglecting contributions from the bottom and charm Yukawa
couplings, the renormalisation group equation for the top Yukawa
coupling at one loop is given by~\cite{Lindner:1985uk}
\begin{align}
  \label{eq:RGE-t}
 \frac{h_{t}^{2}}{d \mu}=\frac{c_{t}}{8\pi^{2}}\,h_{t}^{2}\,\left(h_{t}^{2}-\frac{17}{54}\cdot
  \frac{5}{3}\,g_{1}^{2}-\frac{1}{2}\,g_{2}^{2}-\frac{16}{9}\,g_{3}^{2}\right).
\end{align}
A detailed study of the evolution of quantities relevant for
leptogenesis can be found in~\cite{Antusch:2005gp}.

\providecommand{\href}[2]{#2}\begingroup\raggedright\endgroup

\end{document}